\documentstyle[aaspp4]{article}
\lefthead{Elmegreen, Efremov, Pudritz, \& Zinnecker}
\righthead{Observations and Theory of Star Cluster Formation}
\slugcomment{Protostars and Planets IV}
\begin{document}

\title{OBSERVATIONS AND THEORY OF STAR CLUSTER FORMATION }

\author{B. G. Elmegreen\altaffilmark{1},
Y. Efremov\altaffilmark{2}, R. E. Pudritz\altaffilmark{3},
H. Zinnecker\altaffilmark{4}
}
\altaffiltext{1}{IBM T.J. Watson Research Center,
P.O. Box 218, Yorktown Heights, NY 10598, USA, bge@watson.ibm.com}
\altaffiltext{2}{Sternberg Astronomical Institute, University of Moscow,
Moscow, Russia, efremov@sai.msu.su}
\altaffiltext{3}{Department of Physics and Astronomy, McMaster University,
Hamilton, Ontario, Canada, pudritz@physics.mcmaster.ca}
\altaffiltext{4}{Astrophysikalisches Institut Potsdam, Germany,
hzinnecker@aip.de}

\begin{abstract}
Young stars form on a wide range of scales, producing
aggregates and clusters with various degrees of gravitational
self-binding.  The loose aggregates have a hierarchical structure in
both space and time that resembles interstellar turbulence, suggesting
that these stars form in only a few turbulent crossing times with
positions that map out the previous gas distribution.  Dense clusters,
on the other hand, are often well mixed, as if self-gravitational
motion has erased the initial fine structure.  Nevertheless, some of
the youngest dense clusters also show sub-clumping, so it may be that
all stellar clustering is related to turbulence.  Some of the 
densest clusters may also be triggered. 
The evidence for
mass segregation of the stars inside clusters is reviewed, along with
various explanations for this effect.  Other aspects of the theory of
cluster formation are reviewed as well, including many specific
proposals for cluster formation mechanisms.  The conditions for the formation of bound
clusters are discussed.  Critical star formation efficiencies can be as
low as 10\% if the gas removal process is slow and the stars are born
at sub-virial speeds.  Environmental conditions, particularly pressure,
may affect the fraction and masses of clusters that end
up bound.  Globular clusters may form like normal open clusters but in
conditions that prevailed during the formation of the halo and
bulge, or in interacting
and starburst galaxies today.  Various theories for the formation of globular
clusters are summarized.
\end{abstract}

To be published in Protostars and Planets IV,
eds. V. G. Mannings, A. P. Boss, and S. S. Russell, from
the conference at Santa Barbara, CA, July 6-11, 1998.

\vspace{0.25in}
\centerline{I.~~INTRODUCTION}

The advent of large array cameras at visible to infrared wavelengths,
and the growing capability to conduct deep surveys with semi-automated
searching and analysis techniques, have led to a resurgence in the study
of stellar clusters and groupings in the disk and halo of our Galaxy, in
nearby galaxies, and in distant galaxies. The complementary aspect of
the cluster formation problem, namely the structure of molecular clouds
and complexes, is also being realized by submm continuum mapping and
comprehensive mm surveys. Here we review various theories about the
origin of star clusters and the implications of young stellar clustering
in general, and we discuss the requirements for gravitational
self-binding in open and globular clusters.

Previous reviews of cluster formation were in Wilking \& Lada (1985),
Larson (1990), Lada (1993), Lada, Strom, \& Myers (1993), and Zinnecker,
McCaughrean, \& Wilking (1993).

\vspace{.25in}
\centerline{II.~~STRUCTURE OF YOUNG STELLAR GROUPS}

Stellar groupings basically come in two types: bound and unbound. Some
of the unbound groups could be loose aggregates of stars that formed in
the dense cores of weakly bound or unbound cloud structures, such as
the Taurus star-forming complex.  Other unbound groups could be
dispersed remnants of inefficient star formation in strongly
self-gravitating clouds or cloud cores. 

This section discusses loose aggregates of young stars first, considering
hierarchical structure and a size-time correlation, and then more
compact groups of stars, i.e. cluster formation in dense cloud cores,
along with the associated process of stellar mass segregation, and the
effects (or lack of effects) of high stellar densities on disks,
binaries, and the stellar initial mass function.

\vspace{0.25in}
\centerline{A.~~ Hierarchical Structure in Clouds and Stellar Groups}

\indent{1.~~ Gas Structure }

Interstellar gas is structured into clouds and clumps on a wide range
of scales, from sub-stellar masses ($10^{-4}$ M$_\odot$) that are
hundredths of a parsec in size, to giant cloud complexes ($10^7$
M$_\odot$) that are as large as the thickness of a galaxy. This
structure is often characterized as consisting of discrete clumps or
clouds with a mass spectrum that is approximately a power law,
$n(M)dM\propto M^{-\alpha}dM$, with $\alpha$ in the range from
$\sim1.5$ to $\sim1.9$ (for the smaller scales, see Heithausen et al.
1998 and Kramer et al. 1998, with a review in Blitz 1993; for the
larger scales, see Solomon et al. 1987, Dickey \& Garwood 1989, and the
review in Elmegreen 1993).

Geometrical properties of the gas structure can also be measured from
power spectra of emission line intensity maps (Stutzki et al. 1998). The
power-law nature of the result implies that the emission intensity is
self-similar over a wide range of scales. Self-similar structure has
also been found on cloud perimeters, where a fractal dimension of
$\sim1.3\pm0.3$ was determined (Beech 1987; Bazell \& D\'esert 1988;
Scalo 1990; Dickman, Horvath, \& Margulis 1990; Falgarone, Phillips, \&
Walker 1991; Zimmermann \& Stutzki 1992, 1993; Vogelaar \& Wakker 1994).
This power-law structure includes gas that is both self-gravitating and
non-self-gravitating, so the origin is not purely gravitational
fragmentation. The most likely source is some combination of turbulence
(see review in Falgarone \& Phillips 1991), agglomeration with
fragmentation (Carlberg \& Pudritz 1990; McLaughlin \& Pudritz 1996),
and self-gravity (de Vega, Sanchez, \& Combes 1996).

Interstellar gas is not always fractal. Shells, filaments, and dense
cores are not fractal in their overall shapes, they are regular and have
characteristic scales. Generally, the structures that form as a result
of specific high pressure events, such as stellar or galactic pressures,
have an overall shape that is defined by that event, while the
structures that are formed as a result of turbulence are hierarchical
and fractal. 

\indent{2.~~ Stellar Clustering }

{\it Stellar} clustering occurs on a wide range of scales like most gas
structures, with a mass distribution function that is about the same as
the clump mass distribution function in the gas. For open clusters, it
is a power law with $\alpha$ in the range from $\sim1.5$ (van den Bergh
\& Lafontaine 1984; Elson \& Fall 1985; Bhatt, Pandey, \& Mahra 1991) to
$\sim2$ (Battinelli et al. 1994; Elmegreen \& Efremov 1997), and for OB
associations, it is a power law with $\alpha\sim1.7-2$, as determined
from the luminosity distribution function of HII regions in galaxies
(Kennicutt, Edgar, \& Hodge 1989; Comeron \& Torra 1996; Feinstein 1997;
Oey \& Clarke 1998).

It is important to recognize that the gas and stars are not just clumped,
as in the old plumb-pudding model (Clark 1965), but they are {\it
hierarchically} clumped, meaning that small pieces of gas, and small
clusters, are usually contained inside larger pieces of gas and larger
clusters, through a wide range of scales. Scalo (1985) reviewed these
observations for the gas, and Elmegreen \& Efremov (1998) reviewed the
corresponding observations for the stars. Figure 1 shows a section of
the Large Magellanic Cloud southwest of 30 Dor. Stellar clusterings are
present on many scales, and many of the large clusters contain smaller
clusters inside of them, down to the limit of resolution.

Stellar clumping on a kiloparsec scale was first studied by Efremov
(1978), who identified ``complexes'' of Cepheid variables, supergiants,
and open clusters. Such complexes trace out the recent $\sim30-50$ million
years of star formation in the largest cloud structures, which are the
HI ``superclouds'' (Elmegreen \& Elmegreen 1983, 1987; Elmegreen 1995b)
and ``giant molecular associations'' (Rand \& Kulkarni 1990) that are
commonly found in spiral arms. The sizes of star complexes and the sizes
of superclouds or GMA's are about the same in each galaxy, increasing
regularly with galaxy size from $\sim300$ pc or less in small galaxies
like the LMC to $\sim600$ pc or more in large galaxies like the Milky
Way (Elmegreen et al. 1996). Each star complex typically contains
several OB associations in addition to the Cepheids, because star
formation continues in each supercloud, first in one center and then in
another, for the whole $\sim30-50$ My. This overall timescale is smaller
in smaller galaxies because the complexes are smaller, as Battinelli \&
Efremov (1999) confirmed for the LMC. 

Star complexes are well studied in the Milky Way and local galaxies (see
reviews in Efremov 1989, 1995). In our Galaxy, they were examined most
recently by Berdnikov \& Efremov (1989, 1993), and Efremov \& Sitnik
(1988). The latter showed that {\it 90\% of the young ($\sim10$ My)
clusters and associations in the Milky Way are united into the same star
complexes that are traced by Cepheids and supergiants}. A similarly high
fraction of hierarchical clustering has been found in M31 (Efremov,
Ivanov, \& Nikolov 1987; Battinelli 1991, 1992; Magnier et al. 1993;
Battinelli, Efremov \& Magnier 1996), M33 (Ivanov 1992), the LMC
(Feitzinger \& Braunsfurth 1984), and many other galaxies (Feitzinger \&
Galinski 1987).

A map showing two levels in the hierarchy of stellar structures along
the southwest portion of the western spiral arm in M31 is shown in
figure 2 (from Battinelli, Efremov \& Magnier 1996). The smaller groups,
which are Orion-type OB associations, are shown with faint outlines
displaced 0.1$^\circ$ to the south of the larger groups for clarity; the
smaller groups are also shown as dots with their proper positions inside
the larger groups. Evidently {\it most of the OB associations are within
the larger groups}.

The oldest star complexes are sheared by differential galactic rotation,
and appear as flocculent spiral arms if there are no strong density
waves (Elmegreen \& Efremov 1996). When there are density waves, the
complexes form in the arm crests, somewhat equally spaced, as a result
of gravitational instabilities (Elmegreen \& Elmegreen 1983; Elmegreen
1994; Rand 1995; Efremov 1998).

\indent{3.~~Hierarchical Clustering of Stars on Smaller Scales}

Hierarchical clustering of stars continues from star complexes to OB
associations, down to very small scales.  Infrared observations reveal
embedded clusters of various sizes and densities in star-forming
regions. Many of these, as discussed in the next section, are extremely
dense and deeply embedded in self-gravitating molecular cores. Others
are more open and clumpy, as if they were simply following the
hierarchical gas distribution around them. A good example of the latter
is in the Lynds 1641 cloud in the Orion association, which has several
aggregates comprised of $10-50$ young stars, plus a dispersed
population throughout the cloud (Strom, Strom \& Merrill 1993; Hodapp
\& Deane 1993; Chen \& Tokunaga 1994; Allen 1995).

The distribution of young, x-ray active stars around the sky (Guillout
et al. 1998) is also irregular and clumpy on a range of scales. The
low-mass young stars seen in these x-ray surveys are no longer confined
to dense cores. Sterzik et al. (1995), Feigelson (1996), Covino et al.
(1997), Neuh\"auser (1997), and Frink et al. (1997, 1998) found that the
low mass membership in small star-forming regions extends far beyond the
previously accepted boundaries.  This is consistent with the hierarchical
clustering model.

\indent{4.~~ Two Examples of Hierarchical Stellar Structure:
Orion and W3}

There are many observations of individual clusters that are part of a
hierarchy on larger scales. The Orion region overall contains at least
5 levels of hierarchical structure. On the largest scale (first level),
there is the so-called local arm, or the Orion-Cygnus spur, which has
only young stars (Efremov 1997) and is therefore a sheared star
formation feature, not a spiral density wave (Elmegreen \& Efremov
1996; compare to the Sgr-Car arm, which also has old stars -- Efremov
1997). The largest local condensation (second level) in the
Orion-Cygnus spur is Gould's Belt, of which Orion OB1 is one of several
similar condensations (third level; P\"oppel 1997). Inside Orion OB1
are four subgroups (fourth level; Blaauw 1964), and the youngest of
them, including the Trapezium cluster, contains substructure too (fifth
level): one region is the BN/KL region, perhaps triggered by theta-1c,
and another is near OMC-1S (Zinnecker, McCaughrean, \&
Wilking 1993). The main Trapezium cluster may have no substructure,
though (Bate, Clarke, \& McCaughrean 1998).

A similar hierarchy with five levels surrounds W3. On the largest scale
is the Perseus spiral arm (first level), which contains several giant
star formation regions separated by 1--2 kpc; the W3 complex is in one
of them, and the NGC 7538 complex is in another. The kpc-scale
condensation surrounding W3 (second level) contains associations Cas
OB6 and Per OB1 (which is below the galactic plane and includes the
double cluster h and $\chi$ Per), and these two associations form a
stellar complex. The association Cas OB8, which includes a compact
group of five clusters (Efremov 1989, Fig. 16 and Table 7 on p.  77)
may also be a member of this complex, as suggested by the distances and
radial velocities.  Cas OB6 is the third level for W3. Cas OB6 consists
of the two main star-forming regions W4 (fourth level) and W5, and W4
has three condensations at the edge of the expanded HII region, in the
associated molecular cloud (Lada et al. 1978). W3 is one of these three
condensations, and therefore represents the fifth level in the
hierarchy. The hierarchy may continue further too, since W3 contains
two apparently separate sites of star formation, W3A and W3B
(Wynn-Williams, Becklin, \& Neugebauer 1972; Normandeau, Taylor, \&
Dewdney 1997).

Most young, embedded clusters resemble Orion and W3 in this respect.
They have some level of current star formation activity, with an age
possibly less than $10^5$ years, and are also part of an older OB
association or other extended star formation up to galactic scales,
with other clusters forming in the dense parts here and there for a
relatively long time.

\indent{5.~~ Cluster Pairs and other Small Scale Structure}

Another way this hierarchy appears is in {\it cluster pairs}. Many
clusters in both the Large Magellanic Cloud (Bhatia \& Hatzidimitriou
1988; Kontizas et al. 1989; Dieball \& Grebel 1998; Vallenari et al.
1998) and Small Magellanic Cloud (Hatzidimitriou \& Bhatia 1990) occur
in distinct pairs with about the same age. Most of these binary clusters
are inside larger groups of clusters and stellar complexes. However, the
clumps of clusters and the clumps of Cepheids in the LMC do not usually
coincide (Efremov, 1989, p. 205; Battinelli \& Efremov, 1999).

Some {\it embedded} clusters also have more structure {\it inside} of
them. For example, star formation in the cloud G 35.20-1.74 has
occurred in several different and independent episodes (Persi et al.
1997), and there is also evidence for non-coeval star formation in NGC
3603, the most massive visible young cluster in the Galaxy (Eisenhauer
et al.  1998). W33 contains three separate centers or sub-clusters of
star formation {\it inside} of it that have not yet merged into a
single cluster (Beck et al. 1998). The same is true in 30 Dor and the
associated cluster NGC 2070 (Seleznev 1995), which appears to have
triggered a second generation of star formation in the adjacent
molecular clouds (Hyland et al. 1992;
Walborn \& Blades 1997; Rubio et al. 1998; Walborn et
al. 1999). Similarly, NGC 3603 has substructure with an age difference
of $\sim10$ My, presumably from triggering too (Brandner et al. 1997).
Lada \& Lada (1995) found eight small subclusters with 10
to 20 stars each in the outer parts of IC 348. Piche (1993) found two
levels of hierarchical structure in NGC 2264: two main clusters with
two subclusters in one and three in the other. The old stellar cluster
M67 still apparently contains clumpy outer structure (Chupina \&
Vereshchagin 1998). Some subclusters can even have slightly different
ages: Strobel (1992) found age substructure in 14 young clusters, and
Elson (1991) found spatial substructure in 18 rich clusters in the
LMC.  

Evidence that subclustering did not occur in dense globular clusters was
recently given by Goodwin (1998), who noted from numerical simulations
that initial substructure in globular clusters would not be completely
erased during the short lifetimes of some of the youngest in the LMC.
Because these young populous clusters appear very smooth, their initial
conditions had to be somewhat smooth and spherical too. 

The similarity between the loose clustering properties of many young
stellar regions and the clumpy structure of weakly self-gravitating gas
appears to be the result of {\it star formation following the gas in
hierarchical clouds that are organized by supersonic turbulence.}
Turbulence also implies motion and, therefore, a size-dependent
crossing time for the gas. We shall see in the next section that this
size-dependent timescale might also apply to the duration of star
formation in a region.

\vspace{0.25in}
\centerline{B.~~Star Formation Time Scales}

The duration of star formation tends to vary with the size $S$ of the
region as something like the crossing time for turbulent motions, i.e.,
increasing about as $S^{0.5}$. This means that star formation in larger
structures takes longer than star formation in sub-regions. A schematic
diagram of this time-size pattern is shown in figure 3. The largest
scale is taken to be that of a flocculent spiral arm, which is typically
$\sim100$ My old, as determined from the pitch angle (Efremov \&
Elmegreen 1998).

This relationship between the duration of star formation and the region
size implies that clusters forming together in small regions will
usually have about the same age, within perhaps a factor of three of the
turbulent crossing time of the small scale, while clusters forming
together in larger regions will have a wider range of ages, proportional
to the crossing time on the larger scale. Figure 4 shows this
relationship for 590 clusters in the LMC (Efremov \& Elmegreen 1998).
Plotted on the ordinate is the average difference in age between all
pairs of clusters whose deprojected spatial separations equal the values
on the abscissa. The average age difference between clusters increases
with their spatial separation. In the figure, the correlation ranges
between $0.02^\circ$ and $1^\circ$ in the LMC, which corresponds to a
spatial scale of 15 to 780 pc. The correlation disappears above
$1^\circ$, perhaps because the largest scale for star formation equals
the Jeans length or the disk thickness. A similar duration-size relation
is also observed within the clumps of clusters in the LMC. Larger clumps
of clusters have larger age dispersions (Battinelli \& Efremov 1999).

The correlations between cluster birth times and spatial scale are
reminiscent of the correlation between internal crossing time and size
in molecular clouds. The crossing time in a molecular cloud or cloud
clump is about the ratio of the radius (half-width at half-maximum size)
to the Gaussian velocity dispersion. The data for several molecular
cloud surveys are shown in figure 5, with different symbols for each
survey. On the top is a plot of the Gaussian linewidth versus size, and
on the bottom is the crossing time versus size. Smaller clouds and
clumps have smaller crossing times, approximately in proportion to size
$S^{0.5}.$ Overlayed on this plot, as large crosses, are the
age-difference versus separation points for LMC clusters, from figure 4.
Evidently, the cluster correlation fits in nicely at the top part of the
molecular cloud crossing time-size relation.

These correlations underscore our perception that {\it both cloud
structure, and at least some stellar clusterings, come from
interstellar gas turbulence.} The cluster age differences also suggest
that {\it star formation is mostly finished in a cloud within only
$\sim2$ to 3 turbulent crossing times,} which is very fast. In fact,
this time is much faster than the magnetic diffusion time through the
bulk of the cloud, which is $\sim10$ crossing times in a uniform medium
with cosmic ray ionization (Shu et al. 1987), and even longer if uv
light can get in (Myers \& Khersonsky 1995), and if the clouds are
clumpy (Elmegreen \& Combes 1992). Thus {\it magnetic diffusion does
not regulate the formation of stellar groups}, it may regulate only
the formation of individual stars, which occurs on much smaller scales
(Shu et al. 1987; but see Nakano 1998).

Star formation in a cluster may begin when the turbulent energy of the
cloud dissipates. This is apparently a rapid process, as indicated by
recent numerical simulations of supersonic MHD turbulence, which show a
dissipation time of only 1--2 internal crossing times (MacLow et al.
1998; Stone, Ostriker, \& Gammie 1998). Most giant molecular clouds have
similar turbulent and magnetic energies (Myers \& Goodman 1988) and they
would be unstable without the turbulent energy (McKee et al. 1993), so
the rapid dissipation of turbulence should lead to a similarly rapid
onset of star formation (e.g., McLaughlin \& Pudritz 1996). The
turbulence has to be replenished for the cloud to last more than several
crossing times.

The observed age-size correlation is significantly different from what
one might expect from simple crossing-time arguments in the absence of
turbulence. If the velocity dispersion is independent of scale, as for
an isothermal fluid without correlated turbulent motions, then the slope
of the age-size correlation would be 1.0, not $\sim0.35$. The
correlation is also not from stochastic self-propagating star formation,
which would imply a diffusion process for the size of a star formation
patch, giving a spatial scale that increases as the square root of time.
In that case the slope on figure 4 would be 2.

The duration-size relation for stellar groupings implies that OB
associations and $10^5$ M$_\odot$ GMC's are not physically significant
scales for star formation, but just regions that are large enough to
have statistically sampled the high mass end of the IMF, and young
enough to have these OB stars still present. Regions with such an age
tend to have a certain size, $\sim100$ pc, from the size-time relation, but
the cloud and star formation processes need not be physically distinct. 

The time-scale versus size correlations for star formation should not have
the same coefficients in front of the power laws for all regions of all
galaxies. This coefficient should scale with the total turbulent ISM
pressure to the inverse $1/4$ power (from the relations $P\sim GM^2/R^4$
and $\Delta v^2\sim GM/\left(5R\right)$ for self-gravitating gas;
Chi\`eze 1987; Elmegreen 1989). Thus regions with pressures higher
than the local value by a factor of $10^2-10^4$ should have
durations of star formation shorter than the local regions by a factor
of $3-10$, for the same spatial scale.  This result corresponds to the
observation for starburst galaxies that the formation time of very dense
clusters, containing the mass equivalent of a whole OB association, is
extraordinarily fast, on the order of $\sim1-3$ My, whereas in our
Galaxy, it takes $\sim10$ My to form an aggregate of this mass.
Similarly, high pressure cores in GMCs (Sect. IIC) should form stars
faster than low pressure regions with a similar mass or size.

There are many observations of the duration of star formation in
various regions, both active and inactive. In the Orion Trapezium
cluster, the age spread for 80\% of the stars is very short, less than
1 My (Prosser et al. 1994), as it is in L1641 (Hodapp \& Deane 1993).
It might be even shorter for a large number (but not necessarily a
large fraction) of stars in NGC 1333 because of the large number of
jets and Herbig-Haro objects that are present today (Bally et al.
1996). In NGC 6531 as well, the age spread is immeasurably small
(Forbes 1996). Other clusters have larger age spreads. Hillenbrand et
al. (1993) found that, while the most massive stars (80 M$_\odot$) in
NGC 6611 (=M16) have a 1 My age spread around a mean age of $\sim2$ My,
there are also pre-main sequence stars and a star of 30 M$_\odot$ with
an age of 6 My. The cluster NGC 1850 in the LMC has an age spread of 2
to 10 My (Caloi \& Cassatella 1998), and in NGC 2004, there are evolved
low mass stars in the midst of less evolved high mass stars (Caloi \&
Cassatella 1995). In NGC 4755, the age spread is 6 to 7 My, based on
the simultaneous presence of both high and low mass star formation
(Sagar \& Cannon 1995). One of the best studied clusters for an age
spread is the Pleiades, where features in the luminosity function
(Belikov et al. 1998) and synthetic HR diagrams (Siess et al. 1997)
suggest continuous star formation over $\sim30$ My when it formed
($\sim100$ My ago). This is much longer than the other age spreads for
small clusters, and may have another explanation, including the
possibility that the Pleiades primordial cloud captured some stars from
a neighboring, slightly older, star-forming region (e.g., Bhatt 1989).
Recall that the age spreads are much larger than several My for whole
OB associations and star complexes, as discussed above.

\vspace{0.25in}
\centerline{C.~~Clusters in Dense Molecular Cores}

\indent{1.~~ Cluster Densities}

Infrared, x-ray, and radio continuum maps reveal dense clusters of
young stars in many nearby GMC cores. Reviews of embedded infrared
clusters, including 3-color JHK images, were written by Lada, Strom \&
Myers (1993) and Zinnecker, McCaughrean \& Wilking (1993).

Most observations of embedded young clusters have been made with JHK
imagery. A list of some of the regions studied is in Table 1. These
clusters typically have radii of $\sim0.1$ pc to several tenths of a
pc, and contain several hundred catalogued stars, making the stellar
densities on the order of several times $10^3$ pc$^{-3}$ or larger. For
example, in the Trapezium cluster, the stellar density is $\sim5000$
stars pc$^{-3}$ (Prosser et al. 1994) or higher (McCaughrean \&
Stauffer 1994), and in Mon R2 it is $\sim9000$ stars pc$^{-3}$
(Carpenter et al. 1997). Perhaps the more distant clusters in this list
are slightly larger, as a result of selection effects. 

Some clusters, like W3, NGC 6334, Mon R2, M17, CMa OB1, S106, and the
maser clusters, contain massive stars, even O-type stars in the
pre-UCHII phase or with HII regions. Others, like rho Oph, contain
primarily low mass stars. Although the mass functions vary a little from
region to region, there is no reason to think at this time that the
spatially averaged IMFs in these clusters are significantly different
from the Salpeter (1955), Scalo (1986), or Kroupa, Tout, \& Gilmore
(1993) functions. Thus the clusters with high mass stars also tend to
have low mass stars (Zinnecker, McCaughrean, \& Wilking 1993), although
not all of the low-mass stars are seen yet, and clusters with primarily
low mass stars are not populous enough to contain a relatively rare
massive star (see review of the IMF in Elmegreen 1998a).

Embedded x-ray clusters have been found in NGC 2024 (Freyberg \& Schmitt
1995), IC348 (Preibisch, Zinnecker, \& Herbig 1996), IC1396 (Schulz,
Bergh\"ofer, \& Zinnecker 1997), and the Mon R2 and Rosette molecular
clouds (Gregorio-Hetem et al. 1998). These show x-ray point sources that
are probably T Tauri stars, some of which are seen optically. The
presence of strong x-rays in dense regions of star formation increases
the ionization fraction over previous estimates based only on cosmic ray
fluxes. At higher ionization fractions, magnetic diffusion takes longer
and this may slow the star formation process. For this reason, Casanova
et al. (1995) and Preibisch et al. (1996) suggested that x-rays from T
Tauri stars lead to self-regulation of the star formation rate in dense
clusters. On the other hand, Nakano (1998) suggests that star formation
occurs quickly, by direct collapse, without any delay from magnetic
diffusion. X-rays can also affect the final accretion phase from the
disk. The X-ray irradiation of protostellar disks can lead to better
coupling between the gas and the magnetic fields, and more efficient
angular momentum losses through hydromagnetic winds (cf. K\"onigl \&
Pudritz 1999). Such a process might increase the efficiency of star
formation. The full implications of x-ray radiation in the cluster
environment are not understood yet.

A stellar density of $10^3$ M$_\odot$ pc$^{-3}$ corresponds to an H$_2$
density of $\sim10^4$ cm$^{-3}$.  Molecular cores with densities of
$10^5$ cm$^{-3}$ or higher (e.g., Lada 1992) can easily make clusters
this dense. Measured star formation efficiencies are typically
10\%-40\% (e.g., see Greene \& Young 1992; Megeath et al. 1996; Tapia
et al.  1996). Gas densities of $\sim10^5$ cm$^{-3}$ also imply
extinctions of $A_V\sim40$ mag on scales of $\sim0.2$ pc, which are
commonly seen in these regions, and they imply masses of $\sim200$
M$_\odot$ and virial velocities of $\sim 1$ km s$^{-1}$, which is the
typical order of magnitude of the gas velocity dispersion of cold
star-forming clouds in the solar neighborhood.  There should be larger
and smaller dense clusters too, of course, not a {\it characteristic}
cluster size that is simply the average value seen locally, because
unbiased surveys, as in the LMC (Bica et al. 1996), show a wide range
of cluster masses with power-law mass functions, i.e., no
characteristic scale (cf. Sect. IIA).

\indent{2. Cluster Effects on Binary Stars and Disks}

The protostellar binary fraction is lower in the Trapezium cluster than
the Tau-Aur region by a factor of $\sim3$ (Petr et al. 1998), and lower
in the Pleiades cluster than in Tau-Aur as well (Bouvier et al. 1997).
Yet the binary frequency in the Trapezium and Pleiades clusters are
comparable to that in the field (Prosser et al. 1994). This observation
suggests that most stars form in dense clusters, and that these clusters
reduce an initially high binary fraction at starbirth (e.g., Kroupa
1995a; Bouvier et al. 1997).

The cluster environment should indeed affect binaries. The density of
$n_{star}=10^3$ stars pc$^{-3}$ in a cloud core of size
$R_{core}\sim0.2$ pc implies that objects with this density will collide
with each other in one crossing time if their cross section is
$\sigma\sim \left(n_{star}R_{core}\right)^{-1} \sim0.005$ pc$^2$, which
corresponds to a physical size of
$10^3-10^4\left(R_{core}(pc)n_{star}/10^3\right)^{-1/2}$ AU. This is the
scale for long-period binary stars.

Another indication that a cluster environment affects binary stars is
that the peak in the separation distribution for binaries is smaller
(90 AU) in the part of the Sco-Cen association that contains early type
stars than it is (215 AU) in the part of the Sco-Cen association that
contains no early type stars (Brandner \& K\"ohler 1998).  This
observation suggests that dissipative interactions leading to tighter
binaries, or perhaps interactions leading to the destruction of loose
binaries, are more important where massive stars form.

Computer simulations of protostellar interactions in dense cluster
environments reproduce some of these observations. Kroupa (1995a) got
the observed period and mass-ratio distributions for field binaries by
following the interactions between 200 binaries in a cluster with an
initial radius of $0.8$ pc. Kroupa (1995b) also got the observed
correlations between eccentricity, mass ratio, and period for field
binaries using the same initial conditions. Kroupa (1995c) predicted
further that interactions will cause stars to be ejected from clusters,
and the binary fraction among these ejected stars will be lower than in
the remaining cluster stars (see also Kroupa 1998). These simulations
assume that all stars begin as binary members and interactions destroy
some of these binaries over time.

Another point of view is that the protostars begin as single objects and
capture each other to form binaries. In this scenario, McDonald \&
Clarke (1995) found that disks around stars aid with the capture
process, and they reproduced the field binary fraction in model clusters
with 4 to 10 stars (see review by Clarke 1996). According to this
simulation, the cluster environment should affect disks too. There are
indeed observations of this nature. Mundy et al. (1995) suggested that
massive disks are relatively rare in the Trapezium cluster, and
N\"urnberger et al. (1997) found that protostellar disk mass decreases
with stellar age in the Lupus young cluster, but not in the Tau-Aug
region, which is less dense. When massive stars are present, as in the
Trapezium cluster, uv radiation can photoionize the neighboring disks,
and this is a type of interaction as well (Johnstone et al. 1998).

\indent{3.~~ Cluster Effects on the IMF?}

The best examples of cluster environmental effects on star formation
have been limited, so far, to binaries and disks. Nevertheless, there
are similar suggestions that the cluster environment can affect the
stellar mass as well, and, in doing so, affect the initial stellar mass
function (e.g. Zinnecker 1986).  For example, computer simulations have
been able to reproduce the IMF for a long time using clump (Silk \&
Takahashi 1979; Murray \& Lin 1996) or protostellar (Price \&
Podsiadlowski 1995; Bonnell et al.  1997) interaction models of various
types. 

There is no direct evidence for IMF variations with cluster
density, however (e.g., see Massey \& Hunter 1998; Luhman \& Rieke
1998). Even in extremely dense globular clusters, the IMF seems normal
at low mass (Cool 1998). This may not be surprising because
protostellar condensations are very small compared to the interstar
separations, even in globular clusters (Aarseth et al.  1988), but the
suggestion that massive stars are made by coalescence of smaller
protostellar clumps continues to surface (see Zinnecker et al.  1993;
Stahler, Palla, \& Ho 1999). 

Another indication that cluster interactions do not affect the stellar
mass comes from the observation by Bouvier et al. (1997) that the
rotation rates of stars in the Pleiades cluster are independent of the
presence of a binary companion. These authors suggest that the rotation
rate is the result of accretion from a disk, and so the observation
implies that disk accretion is not significantly affected by
companions.  Presumably this accretion would be even less affected by
other cluster members, which are more distant than the binary
companions.  Along these lines, Heller (1995) found in computer
simulations that interactions do not destroy protostellar disks,
although they may remove $\sim$half of their mass.

There is a way that could have gone unnoticed 
in which the cluster environment may affect the IMF.  This is in the
reduction of the thermal Jeans mass at the high pressure of a cluster-forming
core. A lower Jeans mass might shift the turnover mass in the IMF to
a lower value in dense clusters than in loose groups (Elmegreen 1997, 1999).

\vspace{0.25in}
\centerline{D.~~ Mass Segregation in Clusters}

One of the more perplexing observations of dense star clusters is the
generally centralized location of the most massive stars. This has been
observed for a long time and is usually obvious to the eye. For young
clusters, it cannot be the result of ``thermalization'' because the
time scale for that process is longer than the age of the cluster
(e.g., Bonnell \& Davies 1998). Thus it is an indication of some
peculiar feature of starbirth.

The observation has been quantified using color gradients in 12
clusters (Sagar \& Bhatt 1989), and by the steepening of the IMF with
radius in several clusters (Pandey, Mahra, \& Sagar 1992), including Tr
14 (Vazquez et al. 1996), the Trapezium in Orion (Jones \& Walker 1988;
Hillenbrand 1997; Hillenbrand \& Hartmann 1998), and, in the LMC, NGC
2157 (Fischer et al. 1998), SL 666, and NGC 2098 (Kontizas et al.
1998). On the other hand, Carpenter et al. (1997) found no evidence
from the IMF for mass segregation in Mon R2 at $M<2$M$_\odot$, but
noted that the most massive star (10 M$_\odot$) is near the center
nevertheless. Raboud \& Mermilliod (1998) found a segregation of the
binary stars and single stars in the Pleiades, with the binaries closer
to the center, presumably because of their greater mass. A related
observation is that intermediate mass stars always seem to have
clusters of low mass stars around them (Testi, Palla, \& Natta 1998),
as if they needed these low mass stars to form by coalescence, as
suggested by these authors.

There are many possible explanations for these effects. The stars near
the center could accrete gas at a higher rate and end up more massive
(Larson 1978, 1982; Zinnecker 1982; Bonnell et al. 1997); they (or
their predecessor clumps) could coalesce more (Larson 1990; Zinnecker
et al. 1993; Stahler, Palla, \& Ho 1999; Bonnell, Bate, \& Zinnecker
1998), or the most massive stars and clumps forming anywhere could
migrate to the center faster because of a greater gas drag (Larson
1990, 1991; Gorti \& Bhatt 1995, 1996; Saiyadpour, Deiss, \& Kegel
1997). A central location for the most massive pieces is also expected
in a hierarchical cloud (Elmegreen 1999).  The centralized location of
binaries could be the result of something different: the preferential
ejection of single stars that have interacted with other cluster stars
(Kroupa 1995c).  The presence of low-mass stars around high-mass stars
could have a different explanation too:  high-mass stars are rare so
low-mass stars are likely to form before a high-mass star appears,
whatever the origin of the IMF.

\vspace{0.25in}
\centerline{{I}{I}{I}.~~CLUSTER FORMATION MODELS}

\vspace{0.25in}
\centerline{A.~~Bound Clusters as Examples of Triggered Star Formation?}

Section IIA considered loose stellar groupings as a possible reflection
of hierarchical cloud structure, possibly derived from turbulent
motions, and it considered dense cluster formation in cloud cores
separately, as if this process were different. In fact the two types of
clusters and the processes that lead to them could be related. Even the
bound clusters, which presumably formed in dense cloud cores, have a
power law mass distribution, and it is very much like the power law for
the associations that make HII regions, so perhaps both loose and dense
clusters get their mass from cloud hierarchical structure. The
difference might be simply that dense clusters form in cloud pieces
that get compressed by an external agent.

There are many young
clusters embedded in cores at the compressed interfaces between
molecular clouds and expanded HII regions, including many of those
listed in Table 1 here, as reviewed in Elmegreen (1998b). For example,
Megeath \& Wilson (1997) recently proposed that the embedded cluster in
NGC 281 was triggered by the HII region from an adjacent, older,
Trapezium-like cluster, and Sugitani et al. (1995) found embedded
clusters inside bright rimmed clouds. Compressive triggering of a
cluster can also occur at the interface between colliding clouds, as
shown by Usami et al. (1995). A case in point is the S255 embedded
cluster (Zinnecker et al. 1993; Howard et al. 1997; Whitworth \& Clarke
1997).

Outside compression aids the formation of clusters in several ways. It
brings the gas together so the stars end up in a dense
cluster, and it also speeds up the star formation processes by
increasing the density. These processes can be independent of the
compression, and the same as in other dense regions that were not
rapidly compressed; the only point is that they operate faster in
compressed gas than in lower density gas. The external pressure may
also prevent or delay the cloud disruption by newborn stars, allowing a
large fraction of the gas to be converted into stars, and thereby
improving the chances that the cluster will end up self-bound (cf. Sect
IV; Elmegreen \& Efremov 1997; Lefloch et al. 1997).

Cloud cores should also be able to achieve high densities on their own,
without direct compression. This might take longer, but the usual
processes of energy dissipation and gravitational contraction can lead
to the same overall core structure as the high pressure from an
external HII region. Heyer, Snell \& Carpenter (1997) discussed
the morphology of dense molecular cores and cluster formation in the
outer Galaxy, showing that new star clusters tend to form primarily in
the self-gravitating, high-pressure knots that occur here and there
amid the more loosely connected network of lower pressure material.
Many of these knots presumably reached their high densities
spontaneously.

\vspace{0.25in}
\centerline{B.~~Spontaneous Models and Large Scale Triggering}

The most recent development in cluster formation models is the direct
computer simulation of interacting protostars and clumps leading to
clump and stellar mass spectra (Klessen et al. 1998). Earlier versions
of this type of problem covered protostellar envelope stripping by
clump collisions (Price \& Podsiadlowski 1995), the general stirring
and cloud support by moving protostars with their winds (Tenorio-Tagle
et al.  1993), and gas removal from protoclusters (Theuns 1990).

The core collapse problem was also considered by Boss (1996) who
simulated the collapse of an oblate cloud, forming a cluster with
$\sim10$ stars. A detailed model of thermal instabilities in a
cloud core, followed by a collapse of the dense fragments into the core
center and their subsequent coalescence, was given by Murray \& Lin
(1996).  Patel \& Pudritz (1994) considered core instability with stars
and gas treated as separate fluids, showing that the colder stellar
fluid destabilized the gaseous fluid.

Myers (1998) considered magnetic processes in dense cores, and showed
that stellar-mass kernels could exist at about the right spacing for
stars in a cluster and not be severely disrupted by magnetic waves.
Whitworth et al. (1998) discussed a similar characteristic core size at
the threshold between strong gravitational heating and grain cooling on
smaller scales, and turbulence heating and molecular line cooling on
larger scales.

Some cluster formation models proposed that molecular clouds are made
when high velocity clouds impact the Galactic disk (Tenorio-Tagle
1981). Edvardsson et al. (1995) based this result on abundance
anomalies in the $\zeta$ Sculptoris cluster. Lepine \& Duvert (1994)
considered the collision model because of the distribution of gas and
star formation in local clusters and OB associations, while Phelps
(1993) referred to the spatial distribution, ages, velocities, and
proper motions of 23 clusters in the Perseus arm. Comeron et al. (1992)
considered the same origin for stars in Gould's Belt based on local
stellar kinematics. For other studies of Gould's Belt kinematics, see
Lindblad et al. (1997) and De Zeeuw et al. (1999).

Other origins for stellar clustering on a large scale include
triggering by spiral density waves, which is reviewed in Elmegreen
(1994, 1995a).  According to this model, Gould's Belt was a
self-gravitating condensation in the Sgr-Carina spiral arm when it
passed us $\sim60$ My ago, and is now in the process of large-scale
dispersal as it enters the interarm region, even though there is
continuing star formation in the Lindblad ring and other disturbed gas
from this condensation (see Elmegreen 1993).

The evolution of a dense molecular core during the formation of its
embedded cluster is unknown. The core could collapse
dynamically while the cluster stars form, giving it a total lifetime
comparable to the core crossing time, or it could be somewhat stable as
the stars form on smaller scales inside of it. Indeed there is direct
evidence for gas collapse onto individual stars in cloud cores
(Mardones et al. 1997; Motte, Andre \& Neri 1998), but not much
evidence for the collapse of whole cores (except perhaps in W49 -- see
Welch et al. 1987; De Pree, Mehringer, \& Goss 1997).  

\vspace{0.25in}
\centerline{{I}{V}.~~CONDITIONS FOR THE FORMATION OF BOUND CLUSTERS}

\vspace{0.25in}
\centerline{A.~~ Critical Efficiencies}

The final state of an embedded cluster of young stars depends on the
efficiency, $\epsilon$, of star formation in that region: i.e., on the
ratio of the final stellar mass to the total mass (stars + gas) in that
part of the cloud. When this ratio is high, the stars have enough mass
to remain gravitationally bound when the residual gas leaves, forming a
bound cluster. When this ratio is low, random stellar motions from the
time of birth disperse the cluster in a few crossing times, following
the expulsion of residual gas. The threshold for self-binding occurs at
a local efficiency of about 50\% (von Hoerner 1968). This result is
most easily seen from the virial theorem $2T+\Omega=0$ and total energy
$E=T+\Omega$ for stellar kinetic and potential energies, $T$ and
$\Omega$. Before the gas expulsion, $E=\Omega_{before}/2<0$ from these
two equations. In the instant after {\it rapid} gas expulsion, the
kinetic energy and radius of the cluster are approximately unchanged
because the stellar motions are at first unaffected, but the potential
energy changes because of the sudden loss of mass (rapid gas expulsion
occurs when the outflowing gas moves significantly faster than the
virial speed of the cloud). To remain bound thereafter, $E$ must remain
less than zero, which means that during the expulsion, the potential
energy can increase by no more than the addition of
$|\Omega_{before}|/2$. Thus immediately after the expulsion of gas, the
potential energy of the cluster, $\Omega_{after}$, has to be less than
half the potential energy before, $\Omega_{before}/2$. Writing
$\Omega_{before}=-\alpha GM_{stars}M_{total}/R$ and
$\Omega_{after}=-\alpha GM_{stars}^2/R$ for the same $\alpha$ and $R$,
we see that this constraint requires $M_{stars}>M_{total}/2$ for
self-binding. Thus the efficiency for star formation,
$M_{stars}/M_{total}$, has to exceed about $1/2$ for a cluster to be
self-bound (see also Mathieu 1983; Elmegreen 1983).

Another way to write this is in terms of the expansion factor for
radius, $R_{final}/R_{before}$, where R$_{final}$ is the cluster radius
after the gas-free cluster readjusts its virial equilibrium.  A cluster
is bound if $R_{final}$ does not become infinite. Hills (1980) derived
$R_{final}/R_{initial}=\epsilon/(2\epsilon-1)$, from which we again
obtain $\epsilon>0.5$ for final self-binding with efficiency
$\epsilon$. Danilov (1987) derived a critical efficiency in terms of
the ratio of cluster radius to cloud radius; this ratio has to be
$<0.2$ for a bound cluster to form.

There can be many modifications to this result, depending on the
specific model of star formation. One important change is to consider
initial stellar motions that are less than their virial speeds in the
potential of the cloud because the cloud is supported by both magnetic
and kinematic energies, whereas the star cluster is supported only by
kinematic energy. This modification was considered by Lada, Margulis,
\& Dearborn (1984), Elmegreen \& Clemens (1985), Pinto (1987), and
Verschueren (1990), who derived a critical efficiency for
isothermal clouds that may be approximated by the expression, $$
2\left(1-\epsilon\right)\ln\left({{\epsilon}\over{1-\epsilon}}\right)
+1+\epsilon = 1.5t^2$$ where $t=a_s/a_{VT}<1$ is the ratio of the
stellar velocity dispersion to the virial.  This expression gives
$\epsilon$ between 0.29 at $t=0$ and 0.5 at $t=1$.  Other cloud
structures gave a similar range for $\epsilon$.  This result is the
critical star formation efficiency for the whole cloud; it assumes that
the stars fall to the center after birth, and have a critical
efficiency for binding in the center equal to the standard value of
0.5.

A related issue is the question of purely gravitational effects that
arise in a cluster-forming core once the stars comprise more than
$\sim30$\% of the gas. In this situation, the stars may be regarded as
a separate (collisionless) "fluid" from the gas. The stability of such
two fluid systems was considered by Jog \& Solomon (1984) and Fridman
\& Polyachenko (1984). The Jeans length for a two-component fluid is
smaller than that for either fluid separately. Dense stellar clusters
might therefore fragment into sub-groups, perhaps accounting for some
of the sub-structure that is observed in young embedded star clusters
(Patel \& Pudritz 1994).

Lada, Margulis \& Dearborn (1984) also considered the implications of
slow gas removal on cluster self-binding. They found that gas removal
on timescales of several cloud crossing times lowers the required
efficiency by about a factor of 2, and when combined with the effect of
slow starbirth velocities, lowers the efficiency by a combined factor
of $\sim4$. For clouds in which stars are born at about half the virial
speed, and in which gas removal takes $\sim4$ crossing times, the
critical efficiency for the formation of a bound cluster may be only
$\sim10$\%.

Another way to lower the critical efficiency is to consider gas drag on
the stars that form. Gas drag removes stellar kinetic energy and causes
the stars to sink to the bottom of the cloud potential well, just like
a low birth velocity. Saiyadpour et al. (1997) found that the critical
efficiency can be only 0.1 in this case. Gas accretion also slows down
protostars and causes them to sink to the center (Bonnell \& Davies
1998).

It follows from these examples that the critical efficiency for
self-binding can be between $\sim0.1$ and 0.5, depending on the details of the
star formation process.

\vspace{0.25in}
\centerline{B.~~ Bound Clusters versus Unbound OB Associations}

The onset of massive star formation should mark the beginning of rapid
cloud dispersal
because
ionizing radiation is much more destructive 
per unit stellar mass than short-lived winds from low-mass
stars.
(e.g., see Whitworth 1979). 
According to Vacca, Garmany \& Shull (1996), the ionizing
photon luminosity scales with stellar mass approximately as $M^4$.  In
that case, the total Lyman continuum luminosity from stars with
luminosities in the range $\log L$ to $\log L+d\log L$ increases
approximately as $L^{0.66}$ for a Salpeter IMF (a Salpeter IMF has a
number of stars in a logarithmic interval, $n\left[M_{star}\right]d\log
M_{star}$, proportional to $M_{star}^{-1.35}d\log M_{star}$).  
Thus the total ionizing
luminosity increases with cluster mass more rapidly than the total
cloud mass, and cloud destruction by ionization follows the onset of
massive star formation.

If massive stars effectively destroy clouds, then the overall efficiency
is likely to be low wherever a lot of massive stars form (unless they
form {\it preferentially} late, as suggested by Herbig (1962), and not
just randomly late). Thus we can explain both the low efficiency and the
unboundedness of an OB association: the destructive nature of O-star
ionization causes both. We can also explain why all open clusters in
normal galaxy disks have small masses, generally less than several times
$10^3$ M$_\odot$ in the catalog of Lynga (1987; e.g., see Battinelli et
al. 1994): low mass star-forming regions are statistically unlikely
to produce massive stars.
Discussions of this point are in Elmegreen (1983), Henning \&
Stecklum (1986), and Pandey et al. (1990). 

The idea that massive stars form late in the development of a cluster
goes back to Herbig (1962) and Iben \& Talbot (1966), with more recent
work by Herbst \& Miller (1982) and Adams, Strom \& Strom (1983).
However, Stahler (1985) suggested the observations have a different
explanation, and the rare massive stars should be later than the more
common low-mass star anyway, for statistical reasons (Schroeder \&
Comins 1988; Elmegreen 1999).

The efficiency of star formation has been estimated for several
embedded clusters, giving values such as 25\% for NGC 6334 (Tapia et
al. 1996), 6-18\% for W3 IRS5 (Megeath et al. 1996), 2.5\% for Serpens
(White et al. 1995), 19\% for NGC 3576 (Persi et al. 1994), and 23\%
for rho Oph (Greene \& Young 1992), to name a few.

\vspace{0.25in}
\centerline{C.~~ Variation in Efficiency with Ambient Pressure}

Variations in the efficiency from region to region could have important
consequences because it might affect the fraction of star formation
going into bound clusters (in addition to the overall star formation
rate per unit gas mass). One consideration is that the efficiency may
increase in regions of high pressure (Elmegreen, Kaufman \& Thomasson
1993; Elmegreen \& Efremov 1997). This is because the virial velocity
of a gravitationally-bound cloud increases with pressure and mass as
$V_{VT}\sim (PM^2)^{1/8}$, as may be determined from the relationships
$V_{VT}^2\sim GM/(5R)$ and $P\sim GM^2/R^4$ for radius $R$. If the
pressure increases and the virial velocity follows, then clouds of a
given mass are harder to destroy with HII regions, which push on
material with a fixed velocity of about $10$ km s$^{-1}$.  In fact, a
high fraction of star formation in starburst galaxies, which generally
have a high pressure, could be in the form of bound clusters (Meurer et
al. 1995).

The lack of expansion of HII regions in virialized clouds with high
velocity dispersions also means that the massive stars will not ionize
much. They will only ionize the relatively small mass of high density
gas initially around them.

We can determine the average pressures in today's globular clusters
from their masses and sizes using the relationship $P\sim GM^2/R^4$.
This gives $P\sim 10^6-10^8$ k$_{\rm B}$ (Harris \& Pudritz 1994;
Elmegreen \& Efremov 1997), which is $10^2-10^4$ times the local total
ISM pressure.  If the pressures of star-forming regions in the Galactic
halo were this high when the globular clusters formed, and the globular
cluster cloud masses were higher than those near OB associations by a
factor of $\sim10$, to account for the higher globular cluster masses,
then the velocity dispersions in globular cluster cores had to be
larger than the velocity dispersion in a local GMC by a factor
$\left(M^2P\right)^{1/8}= \left(10^2\times10^4\right)^{1/8} \sim 5.6.$
This puts the dispersion close to $10$ km s$^{-1}$, making the globular
cluster clouds difficult to disrupt by HII regions.

\vspace{0.25in}
\centerline{{V}~~GLOBULAR CLUSTER FORMATION }

Globular clusters in the halos of galaxies are denser, smoother, and
more massive than open clusters in galactic disks, and the globulars are
also much older, but they have about the same power law mass
distribution function as open clusters at the high mass end, and of
course both are gravitationally bound systems. We are therefore faced
with the challenging question of whether the similarities between these
two types of clusters are more important than their differences. If so,
then they may have nearly the same formation mechanisms, modified in the
case of the globulars by the peculiar conditions in the early Universe.
If the differences are too great for a unified model, then we need a
unique formation theory for globular clusters.

The history of the theory on this topic is almost entirely weighted
toward the latter point of view, because the full mass distribution
function for globular clusters is essentially a Gaussian (when plotted
as linear in number versus logarithm in mass or luminosity; e.g., Harris
\& Racine 1979; Abraham \& van den Bergh 1995), with a {\it
characteristic mass} of several $\times10^5$ M$_\odot$. Nearly all of
the early models have attempted to explain this mass. For example,
Peebles \& Dicke (1968), Peebles (1984), Rosenblatt et al. (1988) and
Padoan et al. (1997) regarded globular clusters as primordial objects
produced by density fluctuations in the expanding Universe. Peebles \&
Dickey (1968) thought the characteristic mass was a Jeans mass. Other
models viewed globulars as secondary objects, formed by thermal
instabilities in cooling halo gas (Fall \& Rees 1985; Murray \& Lin
1992; Vietri \& Pesce 1995) or gravitational instabilities in giant
bubbles (Brown et al. 1995) or the shocked layers between colliding
clouds (Zinnecker \& Palla 1987; 
Shapiro, Clocchiatti, \& Kang (1992);
Kumai et al. 1993; Murray \& Lin 1992).
Schweizer (1987) and Ashman \& Zepf (1992) suggested many globulars
formed during galaxy mergers. This could explain the high specific
frequency of globular clusters (number per unit galaxy luminosity;
Harris \& van den Bergh 1981) in ellipticals compared to spirals if the
ellipticals formed in mergers. However, Forbes et al. (1997) found that
galaxies with high specific frequencies of globular clusters have lower
cluster metallicities, whereas the opposite might be expected in the
merger model. Also, McLaughlin (1999) has suggested that the specific
frequency of globular cluster is the same everywhere when x-ray halo gas
and stellar evolution are included.

There is another point of view if the globular cluster mass
function is not primordial but evolved from an initial power law. This
is a reasonable hypothesis because low mass globulars evaporate and get
dispersed first, depressing an initial power law at low mass to
resemble a Gaussian after a Hubble time (Surdin 1979; Okazaki \& Tosa
1995; Elmegreen \& Efremov 1997). Observations of young globular
clusters, forming in starburst regions, also show a power law
luminosity function with a mixture of ages (Holtzman et al. 1992;
Whitmore \& Schweizer 1995; Meurer et al. 1995; Maoz et al. 1996;
Carlson et al. 1998), and the high mass end of the old globular systems
is nearly a power law too (Harris \& Pudritz 1994; McLaughlin \&
Pudritz 1996; Durrell et al. 1996).

In that case, there is a good possibility that old globular clusters
formed in much the same way as young open clusters, i.e., in dense cores
that are part of a large-scale hierarchical gas structure derived from
cloud collisions (Harris \& Pudritz 1994; McLaughlin \& Pudritz 1996) or
turbulent motions (Elmegreen \& Efremov 1997). Direct observations of
globular cluster luminosity functions at cosmological distances should
be able to tell the difference between formation models with a
characteristic mass and those that are scale free.

Another model for globular cluster formation suggests they are the cores
of former dwarf galaxies (Zinnecker et al. 1988; Freeman 1993),
``eaten'' by the large galaxy during dissipative collisions. The
globulars NGC 6715, Terzan 7, Terzan 8, and Arp 2 that are comoving with
the Sgr dwarf galaxy are possible examples (Ibata et al. 1995; Da Costa
\& Armandroff 1995). Other dwarf galaxies have globular cluster systems
too (Durrell et al. 1996), so the globulars around large galaxies may
not come from the cores of the dwarfs, but from the dwarf globulars
themselves. It remains to be seen whether this formation mechanism can
account for the globular cluster luminosity function. 

\vspace{0.25in}
\centerline{{VI}~~CONCLUSIONS}

1. Loose hierarchical clusters form when the associated gas is only weakly
self-gravitating and clumped in this fashion before the star formation
begins. Dense clusters come from strongly self-gravitating gas, which
may be triggered, and which also may be gravitationally unstable to bulk
collapse.

2. Cluster formation is often quite rapid, requiring only a few internal
crossing times to make most of the stars. This follows from the
relatively small age differences between nearby clusters and from the
hierarchical structure of embedded and young stellar groups. Such
structure would presumably get destroyed by orbital mixing if the region
were much older than a crossing time. 

3. Dense cluster environments seem to affect the formation or destruction
of protostellar disks and binary stars, but not the stellar initial mass
function. 

4. Bound clusters require a relatively high star formation efficiency. This
is not a problem for typically low mass open clusters, but it requires
something special, like a high pressure, for a massive globular cluster.

Acknowledgements:  We would like to thank the conference organizers for
the opportunity to write this review.  Helpful comments on the
manuscript were provided by W. Brandner, D. McLaughlin, and P. Kroupa.

Yuri Efremov appreciates partial support from the Russian Foundation
for Basic Research and the Council for Support of Scientific Schools.
The research of Ralph Pudritz is supported by the National Science and
Engineering Research Council of Canada (NSERC).
Travel for Hans Zinnecker was supported by the Deutsche
Forschungsgemeinschaft (DFG).

\begin{table}
\caption{Embedded Young Clusters}
\begin{tabular}{ll}
Region&Reference\\
\tableline
Rho Oph&Wilking \& Lada 1983; Wilking et al. 1985,1989;\\
&Greene \& Young 1992; Comeron et al. 1993; Barsony et al. 1997\\
R Cor Austr& Wilking et al. 1985; Wilking et al. 1997\\
Serpens&White et al. 1995; Hurt \& Barsony 1996; Giovannetti et al. 1998\\
M17&C. Lada et al. 1991; Hanson et al. 1997; Chini \& Wargau 1998\\
L1630&E. Lada et al. 1991; E. Lada 1992; Li et al. 1997\\
Trapezium OMC2&Ali \& Depoy 1995\\
Mon R2&Carpenter et al. 1997\\
Rosette&Phelps \& Lada 1997\\
NGC 281&Henning et al. 1994; Megeath \& Wilson 1997\\
NGC 1333&Aspin et al. 1994; C. Lada et al. 1996\\
NGC 2264&C. Lada et al. 1993; Piche 1993\\
NGC 2282&Horner et al. 1997\\
NGC 3576&Persi et al. 1994\\
NGC 6334&Tapia et al. 1996\\
IC 348&Lada \& Lada 1995\\
W3 IRS5&Megeath et al. 1996\\
S106&Hodapp \& Rayner 1991\\
S255&Howard et al. 1997\\
S269&Eiroa \& Casali 1995\\
BD 40$^\circ$ 4124&Hillenbrand et al. 1995\\
LkH$\alpha$101&Aspin \& Barsony 1994\\
G35.20-1.74&Persi et al. 1997\\
H$_2$O and OH maser sources&Testi et al. 1994\\
19 IRAS sources&Carpenter et al 1993\\
\end{tabular}
\end{table}

\newpage
\begin{figure}
\vspace {6.0in}
\includegraphics{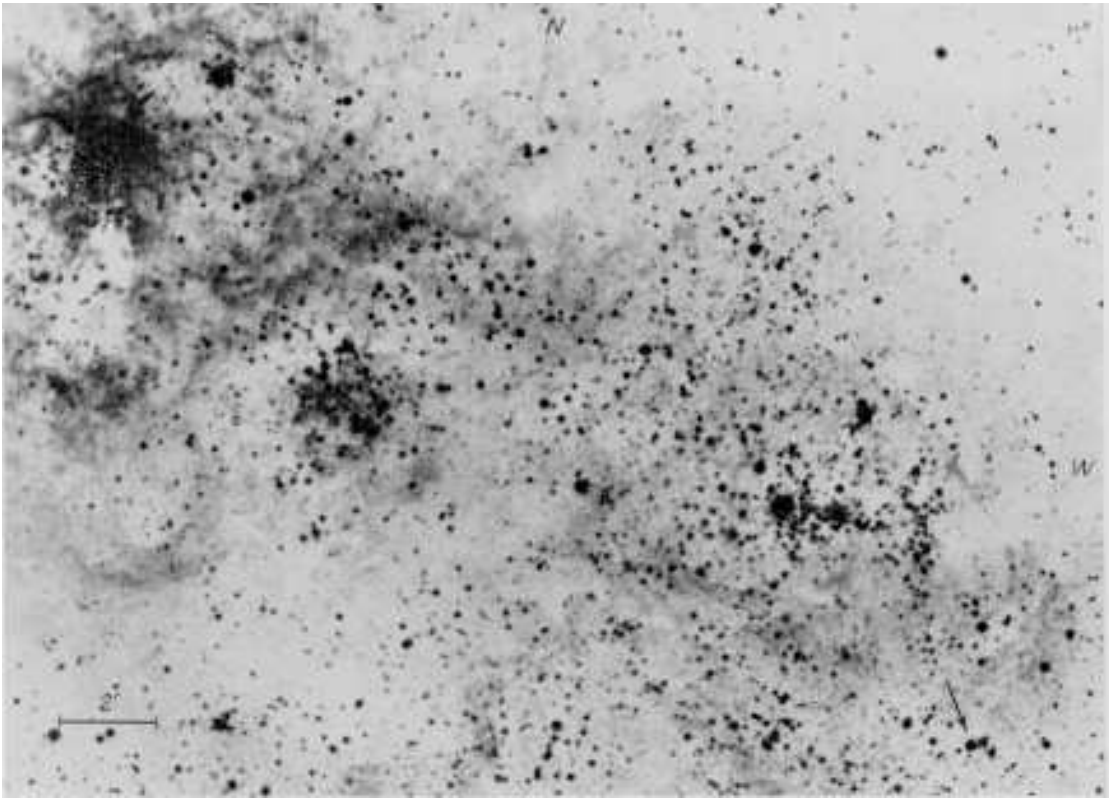}
\caption{Star field southwest of 30 Dor in the Large
Magellanic Clouds, showing hierarchical structure in the stellar
groupings. Image from Efremov (1989).} 
\end{figure}

\newpage
\begin{figure}
\vspace {6.in}
\includegraphics{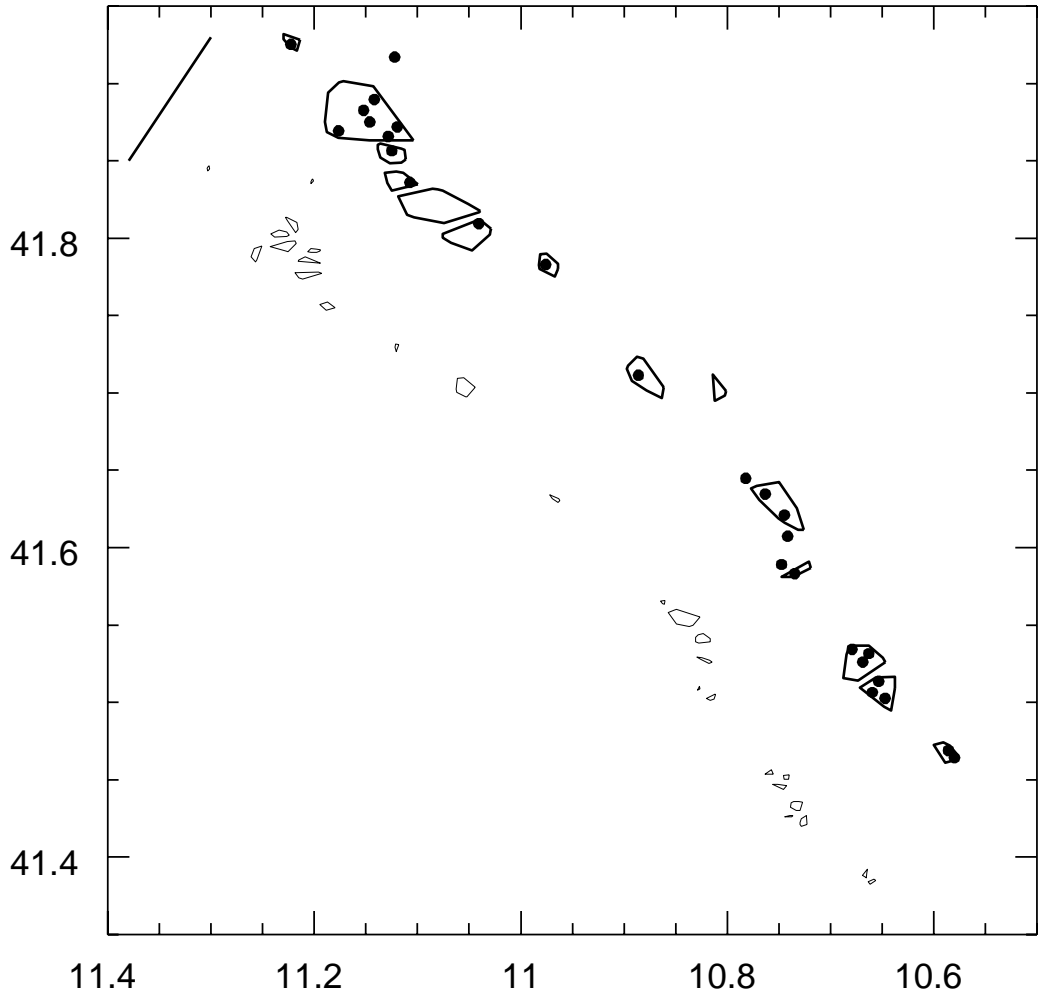}
\caption{OB associations and star complexes along
the western spiral arm of M31. The black dots show the positions of
the OB associations inside the outlines of the star complexes. The
faint outlines to the lower left of this show the actual
OB associations in relation to each other, shifted to the southeast
by the length of the diagonal line for clarity (from Battinelli
et al. 1996).} 
\end{figure}

\newpage
\begin{figure}
\vspace {6.in}
\includegraphics{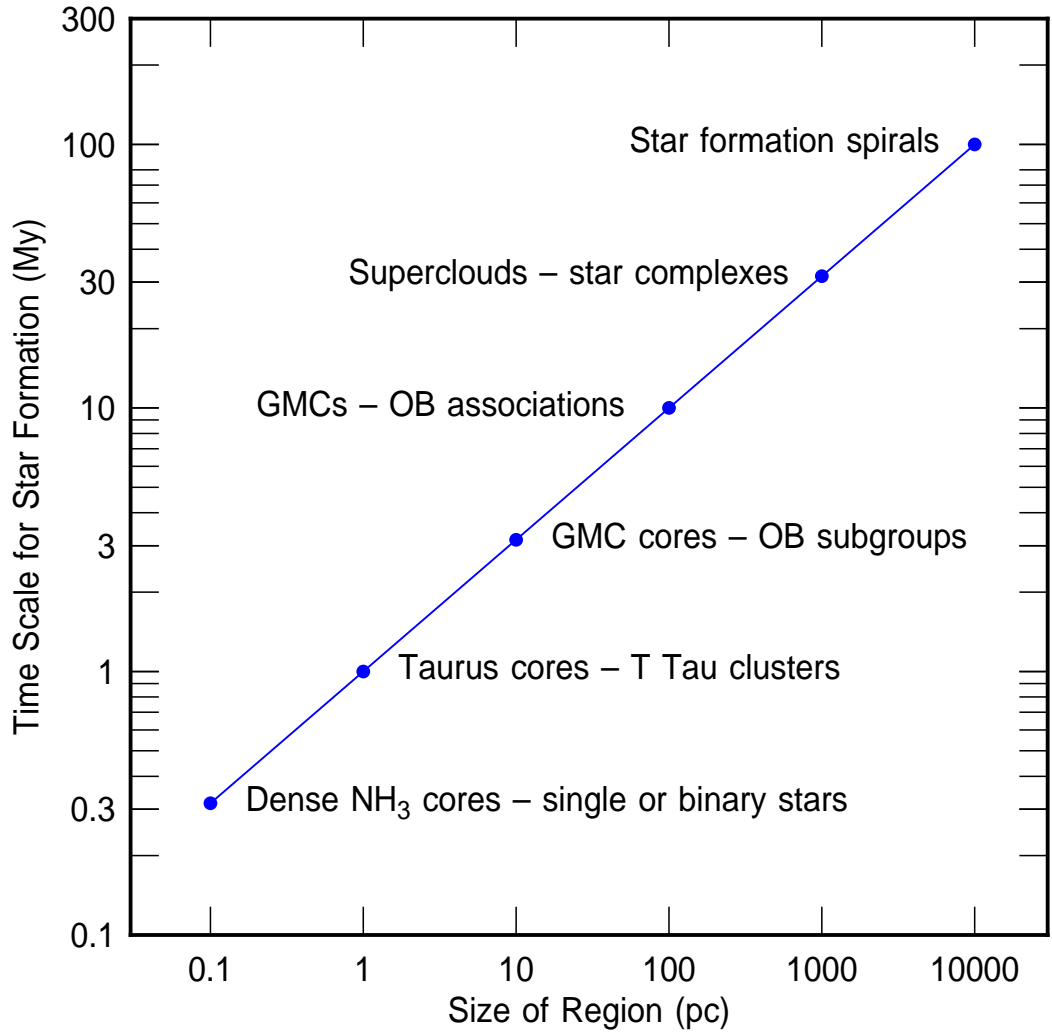}
\caption{Schematic diagram of the relationship
between the duration of star formation and the region size,
from Efremov \& Elmegreen (1998). Larger regions of star
formation form stars for a longer total time. }
\end{figure}

\newpage
\begin{figure}
\vspace {6.in}
\includegraphics{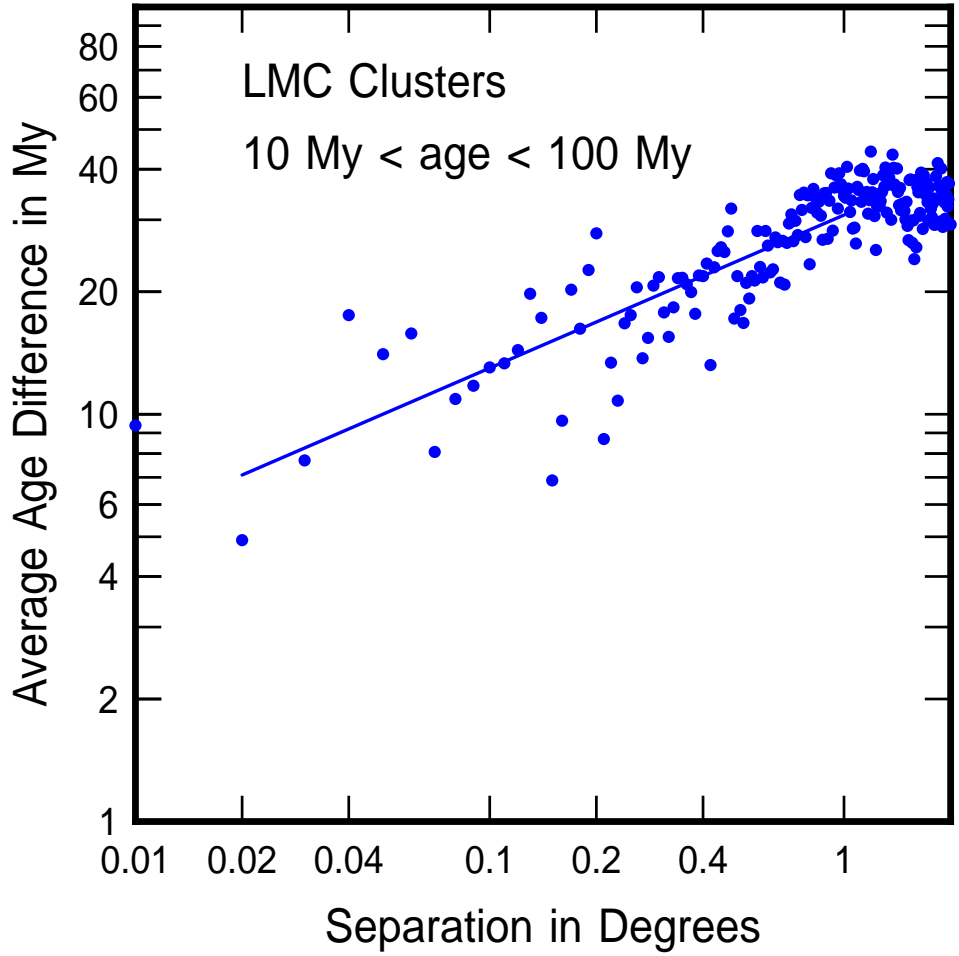}
\caption{The average age differences between clusters  
in the LMC are plotted versus their deprojected angular separations, 
for clusters in the age interval from 10 to 100 My,
from Efremov \& Elmegreen (1998).  Clusters that are close
together in space have similar ages.  The line is a fit
to the data given by $\log \Delta t({\rm yrs})=7.49+0.38\log S({\rm deg})$.}
\end{figure}

\newpage
\begin{figure}
\vspace {6.in}
\includegraphics{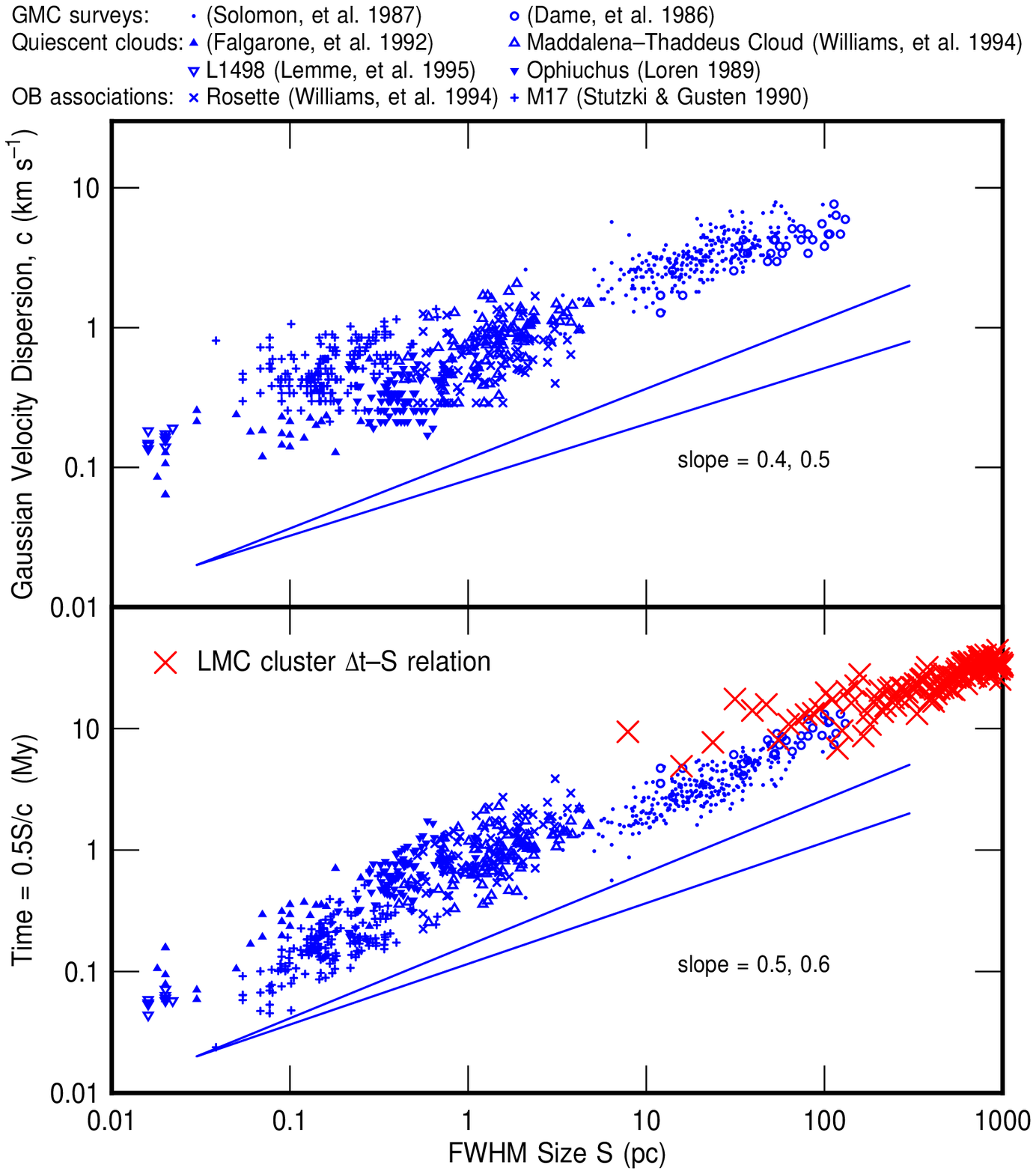}
\caption{The molecular cloud size-linewidth relation
for the Milky Way is shown at the top, considering many different
surveys, as indicated by the symbol types, 
and the ratio of half of the size to the Gaussian
linewidth is shown at the bottom.  This latter ratio
is the crossing time in the cloud; it scales about
as the square root of the cloud size. Superposed on this 
crossing time-size relation is the age-difference versus size relation
shown in the previous figure, using clusters from the LMC. 
If the size-linewidth relations for the two galaxies are
comparable, to within a factor of two, then this diagram suggests that
the duration of star 
formation in a region is approximately equal to the turbulent
crossing time, at least on the large scales considered here
(from Efremov \& Elmegreen 1998). }
\end{figure}


\vskip .5in

\begin{references}
\reference{}{Aarseth, S. J., Lin, D. N. C., and Papaloizou, J. C. B. 1988.
On the collapse and violent relaxation of protoglobular clusters.
{\it ApJ\/} 324:288--310.}

\reference{}{Abraham, R. G., and van den Bergh, S. 1995.
A Gauss-Hermite expansion of the galactic globular cluster luminosity function.
{\it ApJ\/} 438:212--222.}

\reference{}{Adams, M.T., Strom, K.M. and Strom, S.E. 1983.
The star-forming history of the young cluster NGC 2264.
{\it ApJS\/} 53:893--936.}

\reference{}{Allen, L. E. 1995.
Star Formation in Lynds 1641.
PhD Dissertation, Univ. Massachusetts Amherst.}

\reference{}{Ali, B., and Depoy, D. L. 1995.
A 2.2 micrometer imaging survey of the Orion A molecular cloud.
{\it AJ\/} 109:709--720.}

\reference{}{Ashman, K. M., and Zepf, S. E. 1992.
The formation of globular clusters in merging and interacting galaxies.
{\it ApJ\/} 384:50--61.}

\reference{}{Aspin, C., Sandell, G., Russell, A. P. G. 1994.
Near-IR imaging photometry of NGC 1333. I. The embedded PMS stellar population.
{\it A\&A\/} 106:165--198.}

\reference{}{Aspin, C., and Barsony, M. 1994.
Near-IR imaging photometry of the J-K$>$4 
sources in the Lk H{alpha} 101 infrared cluster
{\it A\&A\/} 288:849--859.}

\reference{}{Bally, J., Devine, D., and Reipurth, B. 1996.
A Burst of Herbig-Haro Flows in NGC 1333.
{\it ApJ\/} 473:L49--L53.}

\reference{}{Barsony, M., Kenyon, S. J., Lada, E. A., and Teuben, P. J. 1997.
A Near-Infrared Imaging Survey of the rho Ophiuchi Cloud Core.
{\it ApJS} 112:109-191.}

\reference{}{Bate, M. R., Clarke, C. J., and McCaughrean, M. J. 1998.
Interpreting the mean surface density of companions in star-forming
regions. 
{\it MNRAS\/} 297:1163--1181.}

\reference{}{Battinelli, P. 1991.
A new identification technique for OB associations - OB associations
in the Small Magellanic Cloud.
{\it A\&A\/} 244:69-74.}

\reference{}{Battinelli, P. 1992.
OB associations in four stellar fields of M 31.
{\it A\&A\/} 258:269-271.}

\reference{}{Battinelli P., Brandimarti A., and Capuzzo-Dolcetta R. 1994.
Integrated photometric properties of open clusters.
{\it A\&AS\/} 104:379--390.}

\reference{}{Battinelli, P., Efremov, Y. N., and Magnier, E.A. 1996.
An objective determination of blue star groupings in the Andromeda Galaxy.
{\it A\&A\/} 314:51-58.}

\reference{}{Battinelli, P., and Efremov, Y. N. 1999.
Comparison of distribution of Cepheids and clusters in the LMC.
{\it A\&A\/}, submitted.}

\reference{}{Bazell, D., and D\'esert, F. X. 1988.
Fractal structure of interstellar cirrus.
{\it ApJ\/} 333:353--358.}

\reference{}{Beck, S. C., Kelly, D. M., and Lacy, J. H. 1998.
Infrared Ionic Line Emission in W33.
{\it AJ} 115:2504--2508.}

\reference{}{Beech, M. 1987.
Are Lynds dark clouds fractals?
{\it Ap. Sp. Sci.\/} 133:193--195.}

\reference{}{Belikov, A. N., Hirte, S., Meusinger, H., Piskunov, A. E., and Schilbach, E. 1998.
The fine structure of the Pleiades luminosity function and pre-main sequence evolution.
{\it A\&A\/} 332:575--585.}

\reference{}{Berdnikov L. N., and Efremov, Y. N. 1989.
Groupings of Cepheids in the Galaxy.
{\it  Soviet Ast.} 33:274--279.}  

\reference{}{Berdnikov L. N., and Efremov, Y. N. 1993. 
Contours of constant density and z height for 
Cepheids.
{\it Sov.\ Astr.\ Lett.\/} 19:389--394.}

\reference{}{Bhatia, R.K.,  and  Hatzidimitriou, D. 1988.
Binary star clusters in the Large Magellanic Cloud.
{\it MNRAS\/} 230:215--221.}

\reference{}{Bhatt, H. C. 1989.
Capture of field stars by molecular clouds.
{\it A\&A\/} 213:299--302.}

\reference{}{Bhatt, B.C., Pandey, A.K., and Mahra, H.S. 1991.
Integrated luminosity distribution of galactic open clusters.
{\it J.Ap.Astr.\/} 12:179--185.}

\reference{}{Bica, E., Claria, J. J., Dottori, H., Santos, J. F. C., Jr., and
Piatti, A. E. 1996.
Integrated UBV Photometry of 624 Star Clusters and Associations in
the Large Magellanic Cloud.
{\it ApJS\/} 102:57--73.}

\reference{}{Blaauw, A. 1964.
The O Associations in the Solar Neighborhood.
{\it ARAA\/} 2:213--247.}

\reference{}{Blitz, L. 1993. 
Giant molecular clouds.
In {\it Protostars and Planets III}, eds. E. H. Levy and J. I. Lunine
(Tucson: Univ. Arizona), pp.\ 125--161.}

\reference{}{Bonnell, I. A., Bate, M. R., Clarke, C. J. and Pringle, J. E. 1997.
Accretion and the stellar mass spectrum in small clusters.
{\it MNRAS} 285:201-208.}

\reference{}{Bonnell, I. A., Bate, M. R., and Zinnecker, H. 1998.
On the formation of massive stars.
{\it MNRAS\/} 298:93--102.}

\reference{}{Bonnell, I. A., and Davies, M. B. 1998. 
Mass segregation in young stellar clusters.
{\it MNRAS} 295:691--698}

\reference{}{Boss, A. P. 1996.
Collapse and Fragmentation of Molecular Cloud Cores. IV. Oblate Clouds and Small Cluster
Formation.
{\it ApJ\/} 468:231--240.}

\reference{}{Bouvier, J., Rigaut, F., and Nadeau, D. 1997.
Pleiades low-mass binaries: do companions affect the evolution of protoplanetary disks?
{\it A\&A} 323:139--150.}

\reference{}{Brandner, W., Grebel, E.K., Chu, Y.H., and Weis, K. 1997.
Ring nebula and bipolar outflows associated with the B1.5 supergiant
Sher 25 in NGC 3603.
{\it ApJ\/} 475:L45--L48.}

\reference{}{Brandner, W., and K\"ohler, R. 1998.
Star Formation Environments and the Distribution of Binary Separations. 
{\it ApJ} 499:L79--L82.}

\reference{}{Brown, J. H., Burkert, A., and Truran, J. W. 1995.
On the Formation of Globular Clusters. II. Early Dynamical Evolution.
{\it ApJ\/} 440:666--673.}

\reference{}{Caloi, V., and Cassatella, A. 1995.
On the evolutionary status of the LMC cluster NGC 2004.
{\it A\&A\/} 295:63--74.}

\reference{}{Caloi, V., and Cassatella, A. 1998.
Evolutionary status and age spread in the young LMC cluster NGC 1850A
{\it A\&A\/} 330:492--504.}

\reference{}{Carlberg, R.G. and Pudritz, R., R.E 1990.
Magnetic support and fragmentation of molecular clouds.
{\it MNRAS} 247:353--366.}

\reference{}{Carlson et al. 1998.
Deep Hubble Space Telescope Observations of Star Clusters in NGC 1275.
{\it AJ\/} 115:1778--1790.}

\reference{}{Carpenter, J. M., Snell, R. L., Schloerb, F. P.,
Skrutskie, M. F. 1993.
Embedded star clusters associated with luminous IRAS point sources.
{\it ApJ} 407:657-679.}

\reference{}{Carpenter, J. M., Meyer, M. R., Dougados, C.,
Strom, S. E., Hillenbrand, L. A. 1997.
Properties of the Monoceros R2 Stellar Cluster.
{\it AJ\/} 114:198-221.}

\reference{}{Casanova, S., Montmerle, T., Feigelson, E. D., and Andre, P. 1995.
ROSAT X-ray sources embedded in the rho Ophiuchi cloud core.
{\it ApJ\/} 439:752--770.}

\reference{}{Chen, H., and Tokunaga, A. T. 1994.
Stellar density enhancements associated with IRAS sources in L1641.
{\it ApJS\/} 90:149--172.}

\reference{}{Chi\`eze, J.P. 1987.
The fragmentation of molecular clouds. I - 
The mass-radius-velocity dispersion relations.
{\it A\&A\/} 171:225--232.}

\reference{}{Chini, R., and Wargau, W. F. 1998.
Young stellar objects and abnormal extinction within M17.
{\it A\&A\/} 329:161--168.}

\reference{}{Chupina, N. V., and Vereshchagin, S. V. 1998.
Stellar clumps within the corona in the open cluster M 67.
{\it A\&A\/} 334:552-557.}

\reference{}{Clark, B. G. 1965.
An interferometric investigation of the 21-cm Hydrogen line
absorption. 
{\it Ap.J.\/} 142:1398--1422.}

\reference{}{Clarke, C. J. 1996.
The formation of binaries in small N clusters. 
In {\it The origins, evolution, and destinies of binary stars in clusters,}
ed. E. F. Milone and J.-C. Mermilliod, (San Francisco: ASP) ASP Conf. Ser. 90, pp. 242--251.}

\reference{}{Comeron, F., Torra, J., and Gomez, A. E. 1992.
The characteristics and origin of the Gould's Belt.
{\it ApSpSci\/} 187:187--195.}

\reference{}{Comeron, F., Rieke, G. H., Burrows, A., and Rieke, M. J. 1993.
The Stellar Population in the rho Ophiuchi Cluster.
{\it ApJ\/} 416:185--203.}

\reference{}{Comeron, F., and Torra, J. 1996.
The galactic distribution and luminosity function of ultracompact HII regions.
{\it A\&A\/} 314:776--784.}

\reference{}{Cool, A. M. 1998. 
Measuring globular cluster mass functions with HST.
In {\it The Stellar 
Initial Mass Function}, eds. G. Gilmore,
I. Parry, and S. Ryan, (Cambridge: Cambridge Univ. Press), pp.139--156.} 

\reference{}{Covino, E., Alcala, J.M., Allain, S.,
Bouvier, J., Terranegra,  L., Krautter, J. 1997,
A study of the Chamaeleon star-forming region from the ROSAT all-sky survey.
III. High resolution spectroscopic study
{\it A\&A\/} 328:187--202.}

\reference{}{Da Costa, G. S., and Armandroff, T. E. 1995.
Abundances and Kinematics of the globular cluster systems of the
Galaxy and of the Sagittarius Dwarf.
{\it AJ\/} 109:2533--2552.}

\reference{}{Dame, T., Elmegreen, B.G., Cohen, R., and Thaddeus, P.  1986. 
The largest molecular cloud complexes in the first galactic quadrant.
{\it ApJ} 305:892--908.}

\reference{}{Danilov, V. M. 1987.
The dynamics of forming open stellar clusters.
{\it Astron. Zh.\/} 64:656-659.}

\reference{}{De Pree, C. G., Mehringer, D. M., Goss, W. M. 1997.
Multifrequency, high-resolution radio recombination line observations of the
massive star-forming region W49A.
{\it ApJ\/} 482:307--333.}

\reference{}{de Vega, H., S\'anchez, N., and Combes, F. 1996.
Self gravity as an explanation of the fractal structure of the interstellar medium. 
{\it Nature}, 383:56--58.}

\reference{}{De Zeeuw, P.T., Hoogerwerf, R., de Bruijne, J.H.J., Brown, A.G.A.,
and Blaauw, A. 1999.
A Hipparcos Census of nearby OB associations.
{\it AJ\/}, in press (January 1999).}

\reference{}{Dickey, J. M., and Garwood, R. W. 1989.
The mass spectrum of interstellar clouds.
{\it Ap.J.\/} 341:201--207.}

\reference{}{Dickman, R. L., Horvath, M. A., and Margulis, M. 1990.
A search for scale-dependent morphology in five molecular cloud complexes.
{\it Ap.J\/} 365:586--601.}

\reference{}{Dieball, A., and Grebel, E. 1998.
Binary Star Clusters in the Large Magellanic Cloud.
In {\it IAU Symposium 190, New Views of the Magellanic Clouds},
eds. Yuo-Hua Chu, D.Bohlender, J.Hesser, and N.Suntzeff 
(ASP Conference Series), in press.}

\reference{}{Durrell, P. R., McLaughlin, D. E., Harris, W. E., and Hanes, D. A. 1996.
Globular Cluster Systems in Dwarf Elliptical Galaxies. I. The dE, N Galaxy NGC 3115 DW1.
{\it ApJ\/} 463:543--554.}

\reference{}{Edvardsson, B., Pettersson, B., Kharrazi, M., and Westerlund, B. 1995.
Abundance analysis and origin of the {zeta} Sculptoris open cluster.
{\it A\&A} 293:75--86.}

\reference{}{Efremov, Y. N. 1978.
Star complexes.
{\it Sov.\ Astron.\ Lett.} 4:66.}

\reference{}{Efremov, Y. N. 1989. 
{\it Star Formation Origins in Galaxies} (Moscow: Nauka)}

\reference{}{Efremov, Y. N. 1995.
Star Complexes and Associations: Fundamental and Elementary Cells of Star Formation.
{\it AJ\/} 110:2757--2770.}

\reference{}{Efremov, Y. N. 1997.
Concentration of Cepheids and Open Clusters in the Spiral arms of the Galaxy. 
{\it Astron.Lett.\/} 23:579--584.}

\reference{}{Efremov Yu. N. 1998. 
The Car - Sgr arm as outlined by superclouds
and the grand design of the Galaxy. 
{\it Astron.Ap.Trans.\/} 15:3--17.}

\reference{}{Efremov, Y. N., and Sitnik, T. G. 1988. 
Young stellar-gas complexes in the Galaxy.
{\it Sov.\ Astr.\ Lett.\/} 14:347--352.} 

\reference{}{Efremov, Y. N., Ivanov, G. R., and Nikolov, N. S. 1987.
Star complexes and associations in the Andromeda galaxy.
{\it ApSS\/} 135:119--130.}

\reference{}{Efremov, Y. N., and Elmegreen, B. G. 1998.
Hierarchical Star Formation from 
the Time-Space Distribution of Star Clusters
in the Large Magellanic Cloud.
{\it MNRAS} in press.}

\reference{}{Eiroa, C., and Casali, M. M. 1995.
The S 269 stellar cluster.
{\it A\&A\/} 303:87--94.}

\reference{}{Eisenhauer, F., Quirrenbach, A., Zinnecker, H., and Genzel, R. 1998.
Stellar Content of the Galactic Starburst Template NGC 3603 from
Adaptive Optics Observations. 
{\it ApJ\/} 498:278--292.}

\reference{}{Elmegreen, B. G. 1983. 
Quiescent Formation of Bound Galactic Clusters.
{\it MNRAS\/} 203:1011--1020.}

\reference{}{Elmegreen, B. G. 1989.
A pressure and metallicity dependence for molecular cloud correlations and the
calibration of mass.
{\it ApJ\/} 338:178--196.}

\reference{}{Elmegreen, B. G. 1993. Formation of interstellar clouds and structure.
In {\it Protostars and Planets III}, eds. E. H. Levy and J. I. Lunine
(Tucson: Univ. Arizona), pp.\ 97--124.}

\reference{}{Elmegreen, B. G. 1994.
Supercloud formation by gravitational collapse of magnetic gas in the
crest of a spiral density wave.
{\it ApJ\/} 433:39--47.}

\reference{}{Elmegreen, B. G. 1995a.
Density Waves and Star Formation: Is there Triggering?
In {\it The Formation of the Milky Way},
ed. E.J. Alfaro and A.J. Delgado (Cambridge: Cambridge Univ. Press)
pp. 28--38.}

\reference{}{Elmegreen, B. G. 1995b.
Large Scale Star Formation. In {\it The 7th Guo Shoujing Summer School
on Astrophysics:  Molecular Clouds and Star Formation}, ed.  C. Yuan and
Hunhan You (Singapore: World Press) pp. 149-205.}

\reference{}{Elmegreen, B. G. 1997.
The Initial Stellar Mass Function from Random Sampling in a
Turbulent Fractal Cloud.
{\it ApJ} 486:944--954.}

\reference{}{Elmegreen, B. G. 1998a.
Observations and theory of the initial stellar mass function. 
In {\it Unsolved Problems in Stellar Evolution}, ed. M. Livio, 
(Cambridge: Cambridge Univ. Press), in press.}

\reference{}{Elmegreen, B. G. 1998b.
Observations and Theory of Dynamical Triggers for Star Formation.
In {\it Origins of Galaxies, Stars, Planets and Life},
ed. C.E. Woodward, H.A. Thronson, and M. Shull (San Francisco: 
ASP Conference Series Number 148), pp. 150--183.}

\reference{}{Elmegreen, B. G. 1999.
The initial stellar mass function from random sampling
in hierarchical clouds II: statistical fluctuations and a mass dependence for
starbirth positions and times.
{\it ApJ\/}, 515, in press.}

\reference{}{Elmegreen, B. G., and Elmegreen, D. M. 1983.
Regular strings of H II regions and superclouds in spiral galaxies - Clues to the origin of
cloudy structure
{\it MNRAS} 203:31--45.}

\reference{}{Elmegreen, B. G,. and Clemens, C. 1985.
On the Formation Rate of Galactic
Clusters in Clouds of Various Masses.
{\it ApJ\/} 294:523--532.}

\reference{}{Elmegreen, B. G., and Elmegreen, D. M. 1987.
H I superclouds in the inner Galaxy.
{\it ApJ\/} 320:182--198.}

\reference{}{Elmegreen, B. G., and Combes, F. 1992.
Magnetic diffusion in clumpy molecular clouds.
{\it A\&A\/} 259:232-240.}

\reference{}{Elmegreen, B. G., Kaufman, K., and Thomasson, M. 1993.
An Interaction Model for the Formation of Dwarf Galaxies 
and $10^8$ M$_\odot$ clouds in Spiral Disks. 
{\it ApJ\/} 412:90-98.}

\reference{}{Elmegreen, B.G., Elmegreen, D.M., Salzer, J., and Mann, H. 1996.
On the Size and Formation Mechanism of Star Complexes in Sm, Im, and BCD
Galaxies.
{\it ApJ\/} 467:579--588.}

\reference{}{Elmegreen, B. G., and Efremov, Yu. N. 1996.
An extension of hierarchical star formation to galactic scales.
{\it Ap.J.\/} 466:802--807.}

\reference{}{Elmegreen, B. G., and Efremov, Yu. N. 1997.
A Universal Formation Mechanism for Open and Globular Clusters in Turbulent Gas.
{\it Ap.J.\/} 480:235--245.}

\reference{}{Elmegreen, B. G., and Efremov, Yu. N. 1998.
Hierarchy of Interstellar and Stellar Structures and the
Case of the Orion Star-Forming Region.
In {\it The Orion Complex Revisited}, ed. M. J.
McCaughrean and A. Burkert (San Francisco: ASP Conference Series), 
in press.}

\reference{}{Elson, R. A. W. 1991.
The structure and evolution of rich star clusters in the Large Magellanic Cloud.
{\it ApJS\/} 76:185--214.}

\reference{}{Elson, R.A., and Fall, S.M. 1985.
A luminosity function for star clusters in the Large Magellanic Cloud.
{\it PASP} 97:692--696.}

\reference{}{Falgarone, E., Phillips, T., and Walker, C. K. 1991.
The edges of molecular clouds - Fractal boundaries and density structure.
{\it Ap.J\/} 378:186--201.}

\reference{}{Falgarone, E., and Phillips, T. G. 1991.
Signatures of turbulence in the dense interstellar medium.
In {\it Fragmentation of Molecular Clouds and Star Formation},
eds. E. Falgarone, F. Boulanger, and G. Duvert (Dordrecht: Kluwer) pp. 119--138.}

\reference{}{Falgarone, E., Puget, J.L., and P\'erault, M. 1992. 
The small-scale density and velocity structure of quiescent molecular clouds.
{\it A\&A\/}, 257:715--730.}

\reference{}{Fall, S. M., and Rees, M. J. 1985.
A theory for the origin of globular clusters.
{\it ApJ\/} 298:18--26.}

\reference{}{Feigelson, E. 1996.
Dispersed T Tauri Stars and Galactic Star Formation.
{\it ApJ\/} 468:306--322.}

\reference{}{Feinstein, C. 1997.
H II Regions in Southern Spiral Galaxies: The H alpha Luminosity Function.
{\it ApJS\/} 112:29-47.}

\reference{}{Feitzinger, J. V., and Braunsfurth, E. 1984.
The spatial distribution of young objects in the Large Magellanic Cloud - A problem of
pattern recognition.
{\it A\&A\/} 139:104-114.}

\reference{}{Feitzinger, J. V., and Galinski, T. 1987.
The fractal dimension of star-forming sites in galaxies.
{\it A\&A\/} 179:249--254.}

\reference{}{Fischer, P., Pryor, C., Murray, S., Mateo, M., and Richtler, T. 1998.
Mass Segregation in Young Large Magellanic Cloud Clusters. I. NGC 2157.
{\it AJ\/} 115:592--604.}

\reference{}{Forbes, D. 1996.
Star Formation in NGC 6531-Evidence From the age Spread and Initial Mass Function.
{\it AJ\/} 112:1073--1084.}

\reference{}{Forbes, D. A., Brodie, J. P., and Grillmair, C. J. 1997.
On the Origin of Globular Clusters in Elliptical and cD Galaxies.
{\it AJ\/} 113:1652--1665.}

\reference{}{Freeman, K.H. 1993. 
Globular Clusters and nucleated dwarf ellipticals.
In {\it The Globular Cluster Galaxy Connection},
eds. G.H. Smith and J.P. Brodie (San Francisco: ASP)
ASP Conf. Series, Vol. 48, pp.608--614.}

\reference{}{Freyberg, M. J. and Schmitt, J. H. M. M. 1995.
ROSAT X-ray observations of the stellar clusters in NGC 2023 and NGC 2024.
{\it A\&A\/} 296:L21-L24.}

\reference{}{Fridman, A. M., and Polyachenko, V. L. 1984.
In {\it Physics of Gravitating Systems I. Equilibrium of Stability}
(Berlin: Springer Verlag).}

\reference{}{Frink, S., Roeser, S., Neuh\"auser, R., and Sterzik, M.F. 1997.
New proper motions of pre-main sequence stars in Taurus-Auriga.
{\it A\&A\/} 325:613--622.}

\reference{}{Frink, S. Roeser, S., Alcala, J.M., Covino, E., Brandner, W. 1998.
Kinematics of T Tauri stars in Chamaeleon.
{\it A\&A\/} 338:442--451.}

\reference{}{Giovannetti, P., Caux, E., Nadeau, D., and Monin, J.-L. 1998.
Deep optical and near infrared imaging photometry of the Serpens cloud core.
{\it A\&A\/} 330:990--998.}

\reference{}{Goodwin, S. P. 1998.
Constraints on the initial conditions of globular clusters.
{\it MNRAS} 294:47-60.}

\reference{}{Gorti, U., and Bhatt, H. C. 1995.
Effect of gas drag on the dynamics of protostellar clumps in molecular clouds.
{\it MNRAS\/} 272:61-70.}
                    
\reference{}{Gorti, U., and Bhatt, H. C. 1996.
Dynamics of embedded protostar clusters in clouds.
{\it MNRAS\/} 278:611-616.}

\reference{}{Gregorio-Hetem, J., Montmerle, T., Casanova, S., and Feigelson, E. D. 1998,
X-rays and star formation: ROSAT observations of the Monoceros and Rosette molecular
clouds.
{\it A\&A\/} 331:193-210.}

\reference{}{Greene, T. P., and Young, E. T. 1992.
Near-infrared observations of young stellar objects in the Rho Ophiuchi dark cloud.
{\it ApJ\/} 395:516--528.}

\reference{}{Guillout, P., Sterzik, M. F., Schmitt, J. H. M. M, Motch, C., Egret, D.,
Voges, W., and Neuh\"auser, R. 1998.
The large-scale distribution of X-ray active stars.
{\it A\&A\/} 334:540--544.}

\reference{}{Hanson, M. M., Howarth, I. D., and Conti, P. S. 1997.
The Young Massive Stellar Objects of M17.
{\it ApJ\/} 489:698--718.}

\reference{}{Harris, W. E., and Racine, R. 1979.
Globular clusters in galaxies.
{\it ARAA} 17:241--274.}

\reference{}{Harris, W. E., and van den Bergh, S. 1981.
Globular clusters in galaxies beyond the local group. I - New cluster systems in selected
northern ellipticals.
{\it AJ\/} 86:1627-1642.}

\reference{}{Harris, W. E., and Pudritz, R. E. 1994.
Supergiant molecular clouds and the formation of globular cluster systems.
{\it ApJ\/} 429:177--191.}

\reference{}{Hatzidimitriou, D., and Bhatia, R. K. 1990.
Cluster pairs in the Small Magellanic Cloud.
{\it A\&A\/} 230:11--15.}

\reference{}{Heithausen, A., Bensch, F., Stutzki, J., Falgarone, E.,
and Panis, J.F. 1998.
The IRAM Key Project: small scale structure of pre-star
forming cores. Combined mass spectra and scaling laws. 
{\it A\&A\/} 331:65-68.}

\reference{}{Heller, C. H. 1995.
Encounters with Protostellar Disks. II. Disruption and Binary Formation.
{\it ApJ\/} 455:252--259.}

\reference{}{Henning, Th., and Stecklum, B. 1986.
Self-regulated star formation and the evolution of stellar systems.
{\it ApSpSci\/} 128:237--251.}

\reference{}{Henning, Th., Martin, K., Reimann, H. -G., Launhardt, R.,
Leisawitz, D., and Zinnecker, H. 1994.
Multi-wavelength study of NGC 281 A.
{\it A\&A\/} 288:282--292.}

\reference{}{Herbig, G.H. 1962.
Spectral classification of faint members of the
Hyades and the dating problem in galactic clusters.
{\it ApJ\/} 135:736--747.}

\reference{}{Herbst, W. and Miller, D.P. 1982.
The age spread and initial mass function of NGC 3293:
implications for the formation of clusters.
{\it AJ\/} 87:1478--1490.}

\reference{}{Heyer, M., Snell, R., and Carpenter, J. 1997.
Barometrically Challenged Molecular Clouds.
{\it BAAS\/} 191:121.03.}

\reference{}{Hillenbrand, L. A. 1997.
On the Stellar Population and Star-Forming History of the Orion Nebula Cluster.
{\it AJ} 113:1733--1768.}

\reference{}{Hillenbrand, L. A., Massey, P., Strom, S. E., and
Merrill, K. M. 1993.
NGC 6611: A cluster caught in the act.
{\it AJ\/} 106:1906--1946.}

\reference{}{Hillenbrand, L. A., Meyer, M. R., Strom, S. E.,
Skrutskie, M. F. 1995.
Isolated star-forming regions containing Herbig Ae/Be stars. 1: The young stellar
aggregate associated with BD +40deg 4124.
{\it AJ\/} 109:280-297.}

\reference{}{Hillenbrand, L. A., and Hartmann, L. 1998.
A Preliminary Study of the Orion Nebula Cluster Structure and Dynamics.
{\it ApJ\/} 492:540--553.}

\reference{}{Hills, J. G. 1980.
The effect of mass loss on the dynamical evolution 
of a stellar system - Analytic approximations.
{\it ApJ\/} 235:986--991.}

\reference{}{Hodapp, K.-W. R., and Rayner, J. 1991.
The S106 star-forming region.
{\it AJ} 102:1108-1117.}

\reference{}{Hodapp, K.-W., and Deane, J. 1993.
Star formation in the L1641 North cluster.
{\it ApJS} 88:119-135.}

\reference{}{Holtzman, J. A. et al. 1992.
Planetary Camera observations of NGC 1275 - Discovery of a central population of
compact massive blue star clusters.
{\it AJ\/} 103:691--702.}

\reference{}{Horner, D. J., Lada, E. A., and Lada, C. J. 1997.
A Near-Infrared Imaging Survey of NGC 2282.
{\it AJ\/} 113:1788--1798.}

\reference{}{Howard, E. M., Pipher, J. L., and Forrest, W. J. 1997.
S255-2: The Formation of a Stellar Cluster.
{\it ApJ} 481:327-342.}

\reference{}{Hurt, R. L., and Barsony, M. 1996.
A Cluster of Class 0 Protostars in Serpens: An IRAS HIRES Study.
{\it ApJ} 460:L45--L48.}

\reference{}{Hyland, A. R., Straw, S., Jones, T. J., and Gatley, I. 1992.
Star formation in the Magellanic Clouds. IV - Protostars in the vicinity of 30 Doradus.
{\it MNRAS\/} 257:391--403.}

\reference{}{Ibata, R. A., Gilmore, G., and Irwin, M. J. 1995.
Sagittarius: the nearest dwarf galaxy.
{\it MNRAS\/} 277:781--800.}

\reference{}{Iben I., Jr. and Talbot, R.J. 1966.
Stellar formation rates in young clusters.
\it{ ApJ\/} 144:968--977.}

\reference{}{Ivanov, G. R. 1992.
Stellar associations in M81.
{\it MNRAS} 257:119--124.}

\reference{}{Jog, C. J., and Solomon, P. M. 1984.
Two-fluid gravitational instabilities in a galactic disk.
{\it ApJ\/} 276:114--126.}

\reference{}{Johnstone, D., Hollenbach, D., and Bally, J. 1998.
Photoevaporation of Disks and Clumps by Nearby Massive Stars:
Application to Disk Destruction in the Orion Nebula.
{\it ApJ\/} 499:758--776.}

\reference{}{Jones, B. F., and Walker, M. F. 1988.
Proper motions and variabilities of stars near the Orion Nebula.
{\it AJ\/} 95:1755--1782.}

\reference{}{Kennicutt, R. C., Edgar, B. K., and Hodge, P. W. 1989.
Properties of H II region populations in galaxies. II - The H II region luminosity function.
{\it ApJ} 337:761--781.}

\reference{}{Klessen, R. S., Burkert, A., and Bate, M. R. 1998.
Fragmentation of Molecular Clouds: The Initial Phase of a Stellar Cluster.
{\it ApJ\/} 501:L205-L208.}

\reference{}{K\"onigl, A. and Pudritz,  R.E. 1999. 
Disk Winds and the Accretion-Outflow Connection.
In {\it Protostars and Planets IV},
ed. A. P. Boss, S. S. Russell, and V. Mannings,
(Tucson: Univ. Arizona Press), in press.} 

\reference{}{Kontizas, E., Xiradaki, E., and Kontizas, M. 1989.
The stellar content of binary star clusters in the LMC.
{\it Ap.Sp.Sci.\/} 156:81--84.}

\reference{}{Kontizas, M., Hatzidimitriou, D., Bellas-Velidis, I., Gouliermis, D.,
Kontizas, E., and Cannon, R. D. 1998.
Mass segregation in two young clusters in the Large Magellanic Cloud: SL 666 and NGC
2098.
{\it A\&A\/} 336:503--517.}

\reference{}{Kramer, C., Stutzki, J., R\"ohrig, R., and
Corneliussen, U. 1998.
Clump mass spectra of molecular clouds.
{\it A\&A\/} 329:249--264.}

\reference{}{Kroupa, P. 1995a.
Inverse dynamical population synthesis and star formation.
{\it MNRAS} 277:1491--1506.}

\reference{}{Kroupa, P. 1995b.
The dynamical properties of stellar systems in the Galactic disc.
{\it MNRAS} 277:1507--1521.}

\reference{}{Kroupa, P. 1995c.
Star cluster evolution, dynamical age estimation and the kinematical
signature of star formation.
{\it MNRAS\/} 277:1522--1540.}

\reference{}{Kroupa, P. 1998.
On the binary properties and the spatial and kinematical distribution of
young stars.
{\it MNRAS\/} 298:231--242.}

\reference{}{Kroupa, P., Tout, C.A., and Gilmore, G. 1993.
The distribution of low-mass stars in the Galactic disc.
{\it MNRAS\/}, 262:545--587.}

\reference{}{Kumai, Y., Basu, B., and Fujimoto, M. 1993.
Formation of globular clusters from gas in large-scale unorganized motion in galaxies.
{\it ApJ\/} 404:144--161.}

\reference{}{Lada, C.J. 1993.
The Formation of Low Mass Stars: Observations.
in {\it The Physics of Star Formation and Early
Stellar Evolution}, eds. C. J. Lada and N. D. Kylafis,
Dordrecht:Kluwer, p. 329.}

\reference{}{Lada, C.J., Elmegreen, B.G., Cong, H. and Thaddeus, P. 1978.
Molecular Clouds in the Vicinity of W3, W4, and W5.
{\it ApJ\/} 226:L39--L42.}

\reference{}{Lada, C. J., Margulis, M., and Dearborn, D. 1984.
The formation and early dynamical evolution of bound stellar systems.
{\it ApJ\/} 285:141--152.}

\reference{}{Lada, C. J., Depoy, D. L., Merrill, K. M., Gatley, I. 1991.
Infrared images of M17.
{\it ApJ\/} 374:533--539.}

\reference{}{Lada, C. J., Young, E. T., and Greene, T. P. 1993.
Infrared images of the young cluster NGC 2264.
{\it ApJ\/} 408:471--483.}

\reference{}{Lada, C. J., Alves, J., and Lada, E. A. 1996.
Near-Infrared Imaging of Embedded Clusters: NGC 1333.
{\it AJ\/} 111:1964--1976.}

\reference{}{Lada, E. A., Evans, N. J., II, Depoy D. L., Gatley, I. 1991.
A 2.2 micron survey in the L1630 molecular cloud.
{\it ApJ\/} 371:171--182.}

\reference{}{Lada, E. A. 1992.
Global star formation in the L1630 molecular cloud.
{\it ApJ} 393:L25--L28.}

\reference{}{Lada, E. A.,  Strom, K. M., and Myers, P. C. 1993.
Environments of star formation - Relationship between molecular clouds, dense cores and
young stars.
In {\it Protostars and Planets III}, eds. E. H. Levy and J. I. Lunine
(Tucson: Univ. Arizona), pp.\ 245--277.}

\reference{}{Lada, E. A., and Lada, C. J. 1995.
Near-infrared images of IC 348 and the luminosity functions of young embedded star
clusters.
{\it AJ\/} 109:1682--1696.}

\reference{}{\reference{}{Larson, R. B. 1978. 
Calculations of three-dimensional collapse and fragmentation.
{\it MNRAS\/} 184:69--85.}

\reference{}{Larson, R. B. 1982.
Mass spectra of young stars.
{\it MNRAS\/} 200:159--174.}

\reference{}{Larson, R. B. 1990.
Formation of Star Clusters. In {\it Physical processes in fragmentation and star formation},
eds. . R. Capuzzo-Dolcetta, C.
Chiosi and A. Di Fazio (Dordrecht: Kluwer), pp.\ 389--399.}

\reference{} Larson, R. B. 1991.
Some processes influencing the stellar initial mass function.
In {\it Fragmentation of Molecular Clouds and Star Formation} (ed.
E. Falgarone, F. Boulanger, and G. Duvert). p. 261. Dordrecht: Kluwer. }

\reference{}{Lefloch, B., Lazarell, B., and Castets, A. 1997.
Cometary globules. III. Triggered star formation in IC 1848.
{\it A\&A\/} 324:249--262.}

\reference{}{Lemme, C., Walmsley, C.M., Wilson, T.L., and Muders, D.  1995. 
A detailed study of an extremely quiescent core: L 1498.
{\it A\&A\/} 302:509--520.}

\reference{}{Lepine, J.R.D., and Duvert, G. 1994.
Star formation by infall of high velocity clouds on the galactic disk.
{\it A\&A\/} 286:60-71.}

\reference{}{Li, W., Evans, N. J., II, and Lada, E. A. 1997.
Looking for Distributed Star Formation in L1630: A Near-Infrared (J, H, K) Survey.
{\it ApJ\/} 488:277--285.}

\reference{}{Lindblad, P. O., Palous, J., Loden, K., and Lindegren, L. 1997.
The kinematics and nature of Gould's belt - a 30 Myr old star forming region.
In {\it Hipparcos}, Proceedings of the ESA Symposium ESA SP-402,
ed. B. Battrick, Scientific Coordination: M.A.C. Perryman and 
P. L. Bernacca, pp.507--511.}

\reference{}{Loren, B. R. 1989.
The cobwebs of Ophiuchus. I - Strands of (C-13)O - The mass distribution.
{\it ApJ\/} 338:902--924.}

\reference{}{Luhman, K. L., and Rieke, G. H. 1998.
The Low-Mass Initial Mass Function in Young Clusters: L1495E.
{\it ApJ\/} 497:354--369.}

\reference{}{Lynga, G. 1987.
Catalogue of open cluster data,
5th edition, Strasbourg, CDS}

\reference{}{Mac Low, M.-M., Klessen, R. S., Burkert, A., and Smith, M. D. 1998.
Kinetic Energy Decay Rates of Supersonic and Super-Alvenic Turbulence
in Star-Forming Clouds. 
{\it Phys. Rev. Lett.\/} 80:2754--2757.}

\reference{}{Magnier E.A., Battinelli P., Lewin W.H.G., Haiman, Z.,
van Paradijs, J., Hasinger, G., Pietsch, W., Supper, R., and
Truemper, J. 1993.
Automated identification of OB associations in M31.
{\it A\&A\/} 278:36--42.}

\reference{}{Maoz, D., Barth, A. J., Sternberg, A., Filippenko, A. V.,
Ho, L. C., Macchetto, F. D., Rix, H. W., and Schneider, D. P. 1996.
Hubble Space Telescope Ultraviolet Images of Five Circumnuclear Star-Forming Rings.
{\it AJ\/} 111:2248--2264.}

\reference{}{Mardones, D., Myers, P. C., Tafalla, M., Wilner, D. J., Bachiller, R., and
Garay, G. 1997.
A Search for Infall Motions toward Nearby Young Stellar Objects.
{\it ApJ\/} 489:719-733.}
  
\reference{}{Massey, P., and Hunter, D.A. 1998.
Star Formation in R136: A Cluster of O3 Stars Revealed by Hubble Space Telescope
Spectroscopy.
{\it ApJ\/} 493:180--194.}

\reference{}{Mathieu, R.D. 1983.  
Dynamical constraints on star formation efficiency.
{\it ApJ\/} 267:L97--:101.}

\reference{}{McCaughrean, M. J. and Stauffer, J. R. 1994.
High resolution near-infrared imaging of the Trapezium: A stellar census.
{\it AJ} 108:1382--1397.}

\reference{}{McDonald, J. M., and Clarke, C. J. 1995.
The effect of star-disc interactions on the binary mass-ratio distribution.
{\it MNRAS} 275:671--684.}

\reference{}{McKee, C.F., Zweibel, E.G., Goodman, A.A., and Heiles, C. 1993.
Magnetic fields in star-forming regions -- theory.
In {\it Protostars and Planets III}, eds. E. H. Levy and J. I. Lunine
(Tucson: Univ. Arizona), pp.\ 327--366.}

\reference{}{McLaughlin, D. E. 1999.
The efficiency of globular cluster formation.
{\it ApJ\/} in press.}

\reference{}{McLaughlin, D. E., and Pudritz, R. E. 1996.
The Formation of Globular Cluster Systems. I. The Luminosity Function.
{\it ApJ\/} 457:578--597.}

\reference{}{Megeath, S. T., Herter, T., Beichman, C., Gautier, N., Hester, J. J.,
Rayner, J., and Shupe, D. 1996.
A dense stellar cluster surrounding W3 IRS 5.
{\it A\&A\/} 307:775--790.}

\reference{}{Megeath, S. T., and Wilson, T. L. 1997.
The NGC 281 west cluster. I. Star formation in photoevaporating clumps.
{\it AJ\/} 114:1106-1120.}

\reference{}{Meurer, G. R., Heckman, T. M., Leitherer, C., Kinney, A.,
Robert, C., and Garnett, D. R. 1995.
Starbursts and Star Clusters in the Ultraviolet.
{\it AJ\/} 110:2665--2691.}

\reference{}{Motte, F., Andre, P., and Neri, R. 1998.
The initial conditions of star formation in the rho Ophiuchi main cloud: wide-field
millimeter continuum mapping.
{\it A\&A\/} 336:150--172.}

\reference{}{Mundy, L. G., Looney, L. W., and Lada, E. A. 1995.
Constraints on Circumstellar Disk Masses in the Trapezium Cluster.
{\it ApJ\/} 452:L137-L140.}

\reference{}{Murray, S. D. and Lin, D. N. C. 1992.
Globular cluster formation - The fossil record.
{\it ApJ\/} 400:265--272.}

\reference{}{Murray, S. D. and Lin, D. N. C. 1996.
Coalescence, Star Formation, and the Cluster Initial Mass Function.
{\it ApJ\/} 467:728--748.}

\reference{}{Myers, P. C. 1998.
Cluster-forming Molecular Cloud Cores.
{\it ApJ\/} 496:L109--L112.}

\reference{}{Myers, P. C., and Goodman, A. A. 1988.
Evidence for magnetic and virial equilibrium in molecular clouds.
{\it ApJ\/} 326:L27--L30.}

\reference{}{Myers, P. C., and Khersonsky, V. K. 1995.
On magnetic turbulence in interstellar clouds.
{\it ApJ\/} 442:186--196.}

\reference{}{Nakano, T. 1998.
Star Formation in Magnetic Clouds.
{\it ApJ\/} 494:587--604.}

\reference{}{Neuh\"auser, R.. 1997.
Low mass pre main sequence stars and their x-ray emission.
{\it Science\/} 276:1363--1370.}

\reference{}{Normandeau, M., Taylor, A. R., and Dewdney, P. E. 1997.
The Dominion Radio Astrophysical Observatory Galactic Plane Survey Pilot Project: The
W3/W4/W5/HB 3 Region.
{\it ApJS\/} 108:279--299.}

\reference{}{N\"urnberger, D., Chini, R., and Zinnecker, H. 1997.
 A 1.3mm dust continuum survey of H{alpha} selected T Tauri stars in Lupus.
{\it A\&A\/} 324:1036--1045.}

\reference{}{Oey, M. S., and Clarke, C. J. 1998.
On the Form of the H II Region Luminosity Function.
{\it AJ\/} 115:1543-1553.}

\reference{}{Okazaki, T., and Tosa, M. 1995.
The evolution of the luminosity function of globular cluster systems.
{\it MNRAS} 274:48--60.}

\reference{}{Padoan, P., Jimenez, R., and Jones, B. 1997.
On star formation in primordial protoglobular clouds.
{\it MNRAS} 285:711--717.}

\reference{}{Pandey, A. K., Paliwal, D. C., and Mahra, H. S. 1990.
Star formation efficiency in clouds of various masses.
{\it ApJ\/} 362:165-167.}

\reference{}{Pandey, A. K., Mahra, H. S., and Sagar, R. 1992.
Effect of mass segregation on mass function of young open clusters.
{\it Astr.Soc.India\/} 20:287--295.}

\reference{}{Patel, K., and Pudritz, R. E. 1994.
The formation of stellar groups and clusters in molecular cloud cores.
{\it ApJ} 424:688--713.}

\reference{}{Peebles, P. J. E. 1984.
Dark matter and the origin of galaxies and globular star clusters.
{\it ApJ\/} 277:470-477.}

\reference{}{Peebles, P. J. E., and Dicke, R. H. 1968.
Origin of the globular star clusters. 
{\it ApJ\/} 154:891--908.}

\reference{}{Persi, P., Roth, M., Tapia, M., Ferrari-Toniolo, M., and Marenzi, A. R. 1994.
The young stellar population associated with the HII region NGC 3576
{\it A\&A\/} 232:474--484.}

\reference{}{Persi, P., Felli, M., Lagage, P. O., Roth, M., and Testi, L. 1997.
Sub-arcsec resolution infrared images of the star forming region G 35.20-1.74.
{\it A\&A\/} 327:299--308.}

\reference{}{Petr, M. G., Coude du Foresto, V.T,
Beckwith, S. V. W., Richichi, A., and McCaughrean, M. J. 1998.
Binary Stars in the Orion Trapezium Cluster Core
{\it ApJ\/} 500:825--837.}

\reference{}{Phelps, R. L. 1993.
Young open clusters as probes of the star formation process.
PhD Dissertation, Boston University.}

\reference{}{Phelps, R. L., and Lada, E. A. 1997.
Spatial Distribution of Embedded Clusters in the Rosette Molecular Cloud: Implications for
Cluster Formation.
{\it ApJ\/} 477:176--182.}

\reference{}{Piche, F. 1993.
A Near-Infrared Survey of the Star Forming Region NGC 2264.
{\it PASP} 105:324--324.}

\reference{}{Pinto, F. 1987.
Bound star clusters from gas clouds with low star formation efficiency.
{\it PASP} 99:1161--1166.}

\reference{}{P\"oppel, W. 1997.  
The Gould Belt system and the local interstellar medium.
{\it Fund. Cosmic. Phys.}, 18:1-272.}

\reference{}{Preibisch, T., Zinnecker, H., and Herbig, G. H. 1996.
ROSAT X-ray observations of the young cluster IC 348.
{\it A\&A\/} 310:456--473.}

\reference{}{Price, N. M., and Podsiadlowski, Ph. 1995.
Dynamical interactions between young stellar objects and a 
collisional model for the origin of
the stellar mass spectrum.
{\it MNRAS\/} 273:1041--1068.}

\reference{}{Prosser, C. F., Stauffer, J. R., Hartmann, L.,
Soderblom, D. R., Jones, B. F., Werner, M. W.,
and McCaughrean, M. J. 1994.
HST photometry of the Trapezium cluster.
{\it ApJ\/} 421:517--541.}

\reference{}{Raboud, D., and Mermilliod, J. C. 1998.
Investigation of the Pleiades cluster. IV. The radial structure.
{\it A\&A\/} 329:101--114.}

\reference{}{Rand, R. J., and Kulkarni, S. 1990.
M51 - Molecular spiral arms, giant molecular associations, and superclouds.
{\it ApJ\/} 349:L43--L46.}

\reference{}{Rand, R.J. 1995.
Berkeley-Illinois-Maryland Array Observations 
of Molecular Spiral Structure in M100 (NGC 4321).
{\it AJ\/} 109:2444--2458.}

\reference{}{Rosenblatt, E. I., Faber, S. M., Blumenthal, G. R. 1988.
Pregalactic formation of globular clusters in cold dark matter.
{\it ApJ\/} 330:191-200.}

\reference{}{Rubio, M., Barba, R. H., Walborn, N. R.; Probst, R. G., Garcia, J.,
and Roth, M. R. 1998.
Infrared Observations of Ongoing Star Formation in the 
30 Doradus Nebula and a Comparison with 
Hubble Space Telescope WFPC 2 Images.
{\it AJ\/}, 116:1708--1718.}

\reference{}{Sagar, R., and Bhatt, H. C. 1989.
Radial distribution of the integrated light and photometric colors in open star clusters.
{\it J. Astrop.Astron.\/} 10:173--182.}

\reference{}{Sagar, R., and Cannon, R. D. 1995.
A deep UBVRI CCD photometric study of the moderately young southern open star cluster
NGC 4755 = {kappa} Crucis.
{\it A\&A\/} 111:75-84.}

\reference{}{Saiyadpour, A., Deiss, B. M., and Kegel, W. H. 1997.
The effect of dynamical friction on a young stellar cluster prior to the gas removal.
{\it A\&A} 322:756--763.}

\reference{}{Salpeter, E. E. 1955.
The luminosity function and stellar evolution.
{\it ApJ\/} 121:161--167.}

\reference{}{Scalo, J.M. 1985. 
Fragmentation and Hierarchical Structure in the
Interstellar Medium.  
In {\it Protostars and Planets II}, eds.
D.C. Black and M.S. Matthews (Tucson: Univ. of Arizona), pp. 201-296.}

\reference{}{Scalo, J. M. 1986.
The stellar initial mass function.
{\it Fundam. Cosmic Physics\/} 11:1--278.}
 
\reference{}{Scalo, J. 1990.
Perception of Interstellar Structure: Facing
Complexity.
In {\it Physical Processes in Fragmentation
and Star Formation}  eds.\ R. Capuzzo-Dolcetta, C. Chiosi, and A.
Di Fazio (Dordrecht: Kluwer), pp. 151-178.}  

\reference{}{Schroeder, M. C., and Comins, N. F. 1988.
Star formation in very young galactic clusters.
{\it ApJ\/} 326:756--760.}

\reference{}{Schulz, N. S., Bergh\"ofer, T. W. and Zinnecker, H. 1997.
The X-ray view of the central part of IC 1396.
{\it A\&A\/} 325:1001--1012.}, 

\reference{}{Schweizer F. 1987.
Star formation in colliding and merging galaxies.
In {\it Nearly normal galaxies}, ed. 
S. M. Faber, (New York: Springer-Verlag), pp. 18--25.}

\reference{}{Seleznev, A. F. 1995.
The structure of the halo of the star cluster NGC2070.
{\it Astr. Letters} 21:663--669.}

\reference{}{Shapiro, P. R., Clocchiatti, A., \& Kang, H. 1992.
Magnetic fields and radiative shocks in protogalaxies and the
origin of globular clusters.
{\it ApJ\/} 389:269--285.}
 
\reference{}{Shu, F. H., Adams, F. C., and Lizano, S. 1987.
Star formation in molecular clouds - Observation and theory.
{\it ARAA\/} 25:23--81.}

\reference{}{Siess, L., Forestini, M., and Dougados, C. 1997.
Synthetic Hertzsprung-Russell diagrams of open clusters.
{\it A\&A\/} 324:556--565.}

\reference{}{Silk, J., and Takahashi, T. 1979.
A statistical model for the initial stellar mass function.
{\it ApJ\/} 229:242--256.}

\reference{}{Solomon, P.M., Rivolo, A.R., Barrett, J., and Yahil, A. 1987.
Mass, luminosity, and line width relations of Galactic molecular clouds.
{\it ApJ\/} 319:730--741.}

\reference{}{Stahler, S. W. 1985.
The star formation history of very young clusters.
{\it ApJ\/} 293:207--215.}

\reference{}{Stahler, S. W., Palla, F., and Ho., P. T. P. 1999. 
Formation of massive stars.
In {\it Protostars and Planets IV},
ed. A. P. Boss, S. S. Russell, and V. Mannings,
(Tucson: Univ. Arizona Press), in press.} 

\reference{}{Sterzik, M.F., Alcala, J.M., Neuh\"auser, R., and Schmitt, J.H.M.M. 1995.
The spatial distribution of x-ray selected T Tauri stars. I. Orion.
{\it A\&A\/} 297:418--426.}

\reference{}{Stone, J.M., Ostriker, E., and Gammie, C.F. 1998.
Dissipation in Compressible Magnetohydrodynamic Turbulence.
{\it ApJ\/} 508:L99--L102.}

\reference{}{Strobel, A. 1992.
Age subgroups in open clusters.
{\it A\&A\/} 253:374--378.}

\reference{}{Strom, K. M., Strom, S. E., and Merrill, K. M. 1993.
Infrared luminosity functions for the young stellar population associated with the L1641
molecular cloud.
{\it ApJ\/} 412:233--253.}

\reference{}{Stutzki, J., and G\"usten, R. 1990.
High spatial resolution isotopic CO and CS observations of M17 SW - The
clumpy structure of the molecular cloud core.
{\it ApJ\/} 356:513-533.}

\reference{}{Stutzki, J., Bensch, F., 
Heithausen, A., Ossenkopf, V., and Zielinsky, M. 1998.
On the fractal structure of molecular clouds.
{\it A\&A\/} 336:697--720.}

\reference{}{Sugitani, K., Tamura, M., and Ogura, K. 1995.
Young Star Clusters in Bright-rimmed Clouds: Small-Scale Sequential Star Formation?
{\it ApJ\/} 455:L39--L41.}

\reference{}{Surdin, V.G. 1979.
Tidal destruction of globular clusters in the Galaxy.
{\it Sov.Astr} 23:648--653.}

\reference{}{Tapia, M., Persi, P., and Roth, M. 1996.
The embedded stellar population in northern NGC 6334.
{\it A\&A\/} 316:102--110.}

\reference{}{Tenorio-Tagle, G. 1981.
The collision of clouds with the galactic disk.
{\it A\&A\/} 94:338-344.}

\reference{}{Tenorio-Tagle, G., Munoz-Tunon, C., and Cox, D. P. 1993.
On the Formation of Spheroidal Stellar Systems and the Nature of Supersonic Turbulence in
Star-forming Regions.
{\it ApJ\/} 418:767--773.}

\reference{}{Testi, L., Felli, M., Persi, P., and Roth, M. 1994.
Near-infrared images of galactic masers I. Association between infrared sources and masers.
{\it A\&A\/} 288:634--646.}

\reference{}{Testi, L., Palla, F., and Natta, A. 1998.
A search for clustering around Herbig Ae/Be stars
II. Atlas of the observed sources.
{\it A\&AS\/}, in press.}

\reference{}{Theuns, T. 1990.
A combination of SPH and N-body2 for gas dynamics in star clusters.
In {\it ApSpSci\/} 170:221--224.}

\reference{}{Usami, M., Hanawa, T., and Fujimoto, M. 1995.
High-velocity oblique cloud collisions and gravitational instability of a shock-compressed
slab with rotation and velocity shear.
{\it PASJ\/} 47:271--285.}

\reference{}{Vacca, W.D., Garmany, C.D., and Shull, J.M. 1996.
The Lyman-Continuum Fluxes and Stellar Parameters of O and Early B-Type
Stars.
\it{ ApJ\/} 460:914--931.}

\reference{}{Vallenari, A., Bettoni, D., and Chiosi, C. 1998.
Clusters in the west side of the bar of the Large Magellanic Cloud: interacting pairs?
{\it A\&A\/} 331:506-518.}

\reference{}{van den Bergh, S., and Lafontaine, A. 1984.
Luminosity function of the integrated magnitudes of open clusters.
{\it AJ} 89:1822--1824.}

\reference{}{Vazquez, R. A., Baume, G., Feinstein, A., and Prado, P. 1996.
Investigation on the region of the open cluster Tr 14. 
{\it A\&AS\/} 116:75--94.}

\reference{}{Verschueren, W. 1990.
Collapse of young stellar clusters before gas removal.
{\it A\&A\/} 234:156--163.}

\reference{}{Vietri, M. and Pesce, E. 1995.
Yet another theory for the origin of halo globular clusters and spheroid stars.
{\it ApJ\/} 442:618--627.}

\reference{}{Vogelaar, M. G. R., and Wakker, B. P. 1994.
Measuring the fractal structure of interstellar clouds.
{\it A\&A\/} 291:557--568.}  

\reference{}{von Hoerner, S. 1968.
The formation of stars. In
{\it Interstellar Ionized Hydrogen},
ed. Y. Terzian, (New York: Benjamin), pp. 101--170.}

\reference{}{Walborn, N. R., and Blades, J. C. 1997.
Spectral Classification of the 30 Doradus Stellar Populations.
{\it ApJS} 112:457--485.}

\reference{}{Walborn, N. R., Barba, R. H., Brandner, W., Rubio, M.,
Grebel, E. K., and Probst, R.G. 1999.
Some characteristics of current star formation in the 30 Doradus Nebula
revealed by HST/NICMOS.
{\it AJ\/}, in press.}

\reference{}{Welch, W. J., Dreher, J. W., Jackson, S. M., Tereby, S., and
Vogel, S. N. 1987.
Star formation in W49A: gravitational collapse of a molecular cloud core
toward a ring of massive stars.
{\it Science\/} 238:1550--1555.}

\reference{}{White, G. J., Casali, M. M., and Eiroa, C. 1995.
High resolution molecular line observations of the Serpens Nebula.
{\it A\&A\/} 298:594--605.}

\reference{}{Whitmore, B. C., and Schweizer, F. 1995.
Hubble space telescope observations of young star clusters in NGC-4038/4039, 'the
antennae' galaxies.
{\it AJ\/} 109:960--980.}

\reference{}{Whitworth, A. P. 1979.
The erosion and dispersal of massive molecular clouds by young stars.
{\it MNRAS\/} 186:59--67.}

\reference{}{Whitworth, A. P. and Clarke, C. J. 1997.
Cooling behind mildly supersonic shocks in molecular clouds.
{\it MNRAS\/} 291:578--584.}

\reference{}{Whitworth, A. P., Boffin, H. M. J., and Francic, N. 1998.
The Thermodynamics of Dense Cores. In
{\it Star Formation with the Infrared Space Observatory}, eds.
J. Yun and R. Liseau (APS conf. Series 132), pp. 183--188.}

\reference{}{Wilking, B. A. and Lada, C. J. 1983.
The discovery of new embedded sources in the centrally condensed core of the Rho
Ophiuchi dark cloud - The formation of a bound cluster.
{\it ApJ\/} 274:698--716.}

\reference{}{Wilking, B. A. and Lada, C. J. 1985.
The formation of bound stellar clusters.
In {\it Protostars and Planets II}, eds.\ D.C. Black and M.S. Matthews 
(Tucson: Univ. of Arizona), pp.\ 297--319.}

\reference{}{Wilking, B. A., Harvey, P. M., Joy, M., Hyland, A. R., and Jones, T. J. 1985.
Far-infrared observations of young clusters embedded in the R Coronae Australis and
Rho Ophiuchi dark clouds.
{\it ApJ\/} 293:165--177.}

\reference{}{Wilking, B. A., Lada, C. J., and Young, E. T. 1989.
IRAS observations of the Rho Ophiuchi infrared cluster - Spectral energy distributions
and luminosity function.
{\it ApJ\/} 340:823--852.}

\reference{}{Wilking, B. A., McCaughrean, M. J., Burton, M. G.,
Giblin, T., Rayner, J. T., and Zinnecker, H. 1997.
Deep Infrared Imaging of the R Coronae Australis Cloud Core.
{\it AJ\/} 114:2029--2042.}

\reference{}{Williams, J. P., de Geus, E. J. and Blitz, L. 1994.
Determining structure in molecular clouds.
{\it ApJ\/} 428:693-712.}

\reference{}{Wynn-Williams, C.G., Becklin, E.E., and Neugebauer, G. 1972.
Infra-red sources in the H II region W 3.
{\it MNRAS\/} 160:1--14.}

\reference{}{Zimmermann, T., and Stutzki, J.  1992.
The fractal appearance of interstellar clouds. 
{\it Physica A\/} 191:79--84.}  

\reference{}{Zimmermann, T., and Stutzki, J.  1993.
Fractal aspects of interstellar clouds.
{\it Fractals\/} 1:930--938.}

\reference{}{Zinnecker, H. 1982.
Prediction of the protostellar mass spectrum in the 
Orion near-infrared cluster.
In {\it Symposium on the Orion Nebula to honor
Henry Draper}, eds. A. E. Glassgold, P. J. Huggins, and
E. L. Schucking (New York: New York Academy of Science) pp.226--235.}

\reference{}{Zinnecker, H. 1986. 
IMF in starburst regions.  
In {\it Light on Dark Matter}, ed. F.P. Israel,
ApSS Library Vol. 124, pp.277--278.}

\reference{}{Zinnecker, H., Keable, C. J., Dunlop, J. S., Cannon, R. D.,
and Griffiths, W.K. 1988. 
The nuclei of nucleated dwarf elliptical galaxies - are
they globular clusters?
In {\it The Harlow-Shapley Symposium on Globular
Cluster Systems in Galaxies}, IAU-Symp. 126,
eds. J.E. Grindley and A.G. Davis Philip,
(Dordrecht: Kluwer), pp. 603--604.}

\reference{}{Zinnecker, H., McCaughrean, M. J., and Wilking, B. A. 1993.
The initial stellar population. 
In {\it Protostars and Planets III}, eds. E. H. Levy and J. I. Lunine
(Tucson: Univ. Arizona), pp.\ 429--495.}

\reference{}{Zinnecker, H. and Palla, F. 1987.
Star formation in proto-globular cluster clouds.
In {\it ESO Workshop on Stellar Evolution and Dynamics in the Outer Halo of the Galaxy},
eds. M. Azzopardi and F. Matteucci, (Garching: ESO), p. 355--361.}

\end{references}
\end{document}